\documentclass[a4paper,12pt]{article}

\usepackage{epsfig,graphicx,amsmath,amssymb}
\usepackage{url}
\usepackage{subcaption}
\usepackage[percent]{overpic}
\usepackage{siunitx}
\usepackage{units}
\usepackage{hyperref}
\usepackage{wrapfig}
\newcommand{\bea}{\begin{eqnarray}}
\newcommand{\eea}{\end{eqnarray}}
\newcommand{\HLbLpi}{\mathrm{HLbL};\pi^0}
\newcommand{\HLbLP}{\mathrm{HLbL; P}}
\newcommand{\HLbLPone}{\mathrm{HLbL; P(1)}}
\newcommand{\HLbLPtwo}{\mathrm{HLbL; P(2)}}
\newcommand{\FFP}{{\cal F}_{{\rm P}\gamma^*\gamma^*}}
\newcommand{\FFa}{{\cal F}_{\pi^0\gamma^*\gamma^*}}
\usepackage{relsize}
\RequirePackage{xspace}
\newcommand{\gev}{\ensuremath{\mathrm{\,Ge\kern -0.1em V}}\xspace}
\def\babar{\mbox{\slshape B\kern-0.1em{\smaller A}\kern-0.1em
    B\kern-0.1em{\smaller A\kern-0.2em R}}}

\def\Y#1S{\ensuremath{\Upsilon{(#1S)}}\xspace}

\begin{document}

\thispagestyle{empty}

$\phantom{.}$

\begin{flushright}
{\sf  MITP/16-086 \\
  } 
\end{flushright}

\hfill

\begin{center}
{\Large {\bf Mini-Proceedings, 18th meeting of the Working Group on Radiative Corrections and MC Generators for Low Energies} \\
\vspace{0.75cm}}

\vspace{1cm}

{\large 19$^{\mathrm{th}}$ - 20$^{\mathrm{st}}$ May, Laboratori Nazionali di Frascati, Italy}

\vspace{2cm}

{\it Editors}\\
Pere Masjuan (Mainz), Graziano Venanzoni (Frascati)
\vspace{2.5cm}

ABSTRACT

\end{center}

\vspace{0.3cm}

\noindent
The mini-proceedings of the 18$^{\mathrm{th}}$ Meeting of the "Working Group on Radiative Corrections and MonteCarlo Generators for Low Energies" held in Frascati, 19$^{\mathrm{th}}$ - 20$^{\mathrm{st}}$ May, are presented. These meetings, started in 2006, have as aim to bring together experimentalists and theoreticians working in the fields of meson transition form factors, hadronic contributions to the anomalous magnetic moment of the leptons, and the effective fine structure constant. The development of MonteCarlo generators and Radiative Corrections for precision $e^+e^-$ and $\tau$-lepton physics are also covered, with emphasis on meson production. At this workshop, a documentary entitled {\it Bruno Touschek with AdA in Orsay} commemorating the first observation of electron-positron collisions in a laboratory was also presented. With this edition, the working group reaches 10 years of continuous activities.

\medskip\noindent
The web page of the conference:
\begin{center}
\url{https://agenda.infn.it/conferenceDisplay.py?ovw=True&confId=11309} 
\end{center}
\noindent
contains the presentations.
\vspace{0.5cm}

\noindent
We acknowledge the support and hospitality of the Laboratori Nazionali di Frascati.

\newpage

{$\phantom{=}$}

\vspace{0.5cm}

\tableofcontents

\newpage

\section{Introduction to the $18^{th}$ Radio MontecarLow Working Group meeting}

\addtocontents{toc}{\hspace{1cm}{\sl G.~Venanzoni}\par}

\vspace{5mm}

\noindent
 H.~Czy\.z$^1$ and G.~Venanzoni$^2$

\vspace{5mm}

\noindent
$^1$Institute of Physics, University of Silesia, 40007 Katowice, Poland\\
$^2$Laboratori Nazionali di Frascati dellÕINFN, 00044 Frascati, Italy

\vspace{5mm}
 
The importance of continuous and close collaboration between the experimental and theoretical groups is crucial in the quest for
precision in hadronic physics. This is the reason why the Working Group on ``Radiative Corrections and Monte Carlo Generators for Low Energies'' (Radio MonteCarLow)  was formed a few years ago bringing together experts (theorists and experimentalists) working in the field of low-energy $e^+e^-$ physics and partly also the $\tau$ community.
Its main motivation was to understand the status and the precision of the Monte Carlo generators (MC) used to analyze the hadronic cross section measurements obtained as well with energy scans as with radiative return, to determine luminosities, and whatever possible to perform tuned comparisons, {\it i.e.} comparisons of MC generators with a common set of input parameters and experimental cuts. This  main effort was summarized in a report published in 2010~\cite{Actis:2010gg1}.
During the years the WG structure has been enriched of more physics items and now it includes seven subgroups: Luminosity, R-measurement, ISR, Hadronic VP $g-2$ and Delta alpha, gamma-gamma physics, FSR models, tau decays. 

The present workshop, being its 18$^{th}$ edition of the Radio MonteCarLow meeting coincides with the commemoration of 10 years of activities of the working group which, starting in 2006 with a first meeting held in Frascati, has gathered theoreticians and experimentalists discussing the precision frontier in hadronic physics. A summary of the last years' activities can be found in Refs.~\cite{Czyz:2013zga,Czyz:2013sga,vanderBij:2014mxa,Carloni:2014mpa,Czyz:2015nna}.

During the workshop the last achievements of each subgroups have been presented.
The present accuracy and the future prospects of MC generators for  $e^+e^-$ into leptonic, $\gamma\gamma$,  hadronic and tau final states have been reviewed. 
Recent proposals on the evaluation of the hadronic light-by-light scattering contribution to $g-2$ of the muon and its relation to data which can be useful to reduce the uncertainty on the model dependence have been discussed. Meson Transition FF and proton FF measurements have been reviewed.
New results from CMD3 have been presented, together with new Monte Carlo generator developments.

At this workshop, we have presented as well a documentary entitled {\it Bruno Touschek with AdA in Orsay}, which was prepared in 2013 to commemorate the first observation of electron-positron collisions in a laboratory. 

The workshop was held from the 19$^{th}$ to the 20$^{st}$ May 2016, at the Laboratori Nazionali di Frascati dell'INFN, Italy.
\\
\\
The Webpage of the conference is 
\begin{center}
\url{https://agenda.infn.it/conferenceDisplay.py?ovw=True&confId=11309} 
\end{center}
\noindent
where the detailed program and talks can be found.
\\
\\
\\
All the information on the WG can be found at the web page:
\begin{center}
\url{http://www.lnf.infn.it/wg/sighad/} 
\end{center}

\newpage

\section{Summaries of the talks: Transition Form Factors and Light by Light}

\subsection{Transition Form Factors of pseudoscalars - current experimental status}
\addtocontents{toc}{\hspace{2cm}{\sl S.~Eidelman}\par}

\vspace{5mm}

S.~Eidelman$^{1,2}$

\vspace{5mm}

\noindent
$^1$~Budker Institute of Nuclear Physics, Novosibirsk 630090, Russia \\
$^2$~Novosibirsk State University, Novosibirsk 630090, Russia \\
\vspace{5mm}

Studies of meson transition form factors (TFF) have recently got a new
powerful impetus after realizing their importance for the muon g-2
anomaly~\cite{Czerwinski:2012ry}. Here we restrict ourselves to a brief 
discussion of the current status of TFF studies in decays of pseudoscalars.

The simplest decay to a lepton pair is helicity suppressed, so the
probability of decay to a lepton pair, $P \to l^+l^-$, is proportional to 
$(m_l/m_P)^2$ making the corresponding branchings very low. In 
Table~\ref{tab1} we describe the status of such decays showing for comparison 
a so called unitarity bound - a minimal possible branching fraction that
corresponds to a real part of the decay amplitude. A non-zero imaginary part 
or effects of New Physics (NP) can significantly enhance it. In this Table 
and throughout this Note we list the most precise measurement only, while
the full status can be found in the PDG listings~\cite{Agashe:2014kda}.  
\begin{center}
\begin{table}[h]
\caption{\label{tab1}Status of  $P \to l^+l^-$ decays}  
\begin{tabular}{|l|c|c|c|c|}
\hline
Decay mode & ${\cal {B}}_{\rm exp}$ & Events & Group & 
${\cal {B}}_{\rm unit. bound}$ \\ 
\hline
$\pi^0 \to e^+e^-$ & $(6.46 \pm 0.33) \cdot 10^{-8}$ & 794  & 
KTEV, 2007~\cite{Abouzaid:2006kk} & $4.8 \cdot 10^{-8}$ \\
\hline
$\eta \to e^+e^-$ & $< 2.3 \cdot 10^{-6}$  & -- & HADES, 2012~\cite{Agakishiev:2013fwl} & 
$1.8 \cdot 10^{-9}$ \\
\hline
$\eta \to \mu^+\mu^-$ & $(5.7 \pm 0.9) \cdot 10^{-6}$  &
 114 & SATURNEII, 1994~\cite{Abegg:1994wx} & $4.3 \cdot 10^{-6}$ \\
\hline
$\eta^\prime \to e^+e^-$ & $< 5.6 \cdot 10^{-9}$ & -- & 
CMD-3/SND, 2015~\cite{Achasov:2015mek} & $3.75 \cdot 10^{-11}$  \\
\hline
\end{tabular}
\end{table}
\end{center}
As shown recently in Refs.~\cite{Akhmetshin:2014hxv,Achasov:2015mek,Achasov:2015ark}, a study of
the inverse reaction $e^+e^- \to P \to f$, where $f$ denotes a
decay mode providing a faborable signal-to-nosie ratio, substantially 
enhances our potential in searches for $\eta$ and $\eta'$ to an $e^+e^-$
pair.  


Dalitz decays in which $q^2_1=0$ and $4m^2_l < q^2_2 < m^2_P$ allow
a real study of the TFF as a function of corresponding $q^2$. The
status of branching fraction measurements is summarized in Table~\ref{tab2}.
Reliable results for TFF and their slopes exist for the $\eta$ meson only.
\begin{center}
\begin{table}[h]
\caption{\label{tab2}Status of  $P \to l^+l^-\gamma$ decays}

\begin{tabular}{|l|c|c|c|c|}
\hline
Decay mode & ${\cal {B}}$ & Events & Group & Process \\
\hline
$\pi^0 \to e^+e^-\gamma$ & $(1.174 \pm 0.035) \cdot 10^{-2}$ & 12k  & 
ALEPH & $e^+e^- \to Z$, 2008~\cite{Beddall:2008zza} \\
\hline
$\eta \to e^+e^-\gamma$ & $(6.9 \pm 0.4) \cdot 10^{-3}$  & 1345 & Cr.Ball 
& $\gamma p \to p \eta$, 2011~\cite{Berghauser:2011zz} \\
\hline
$\eta \to \mu^+\mu^-\gamma$ & $(3.1 \pm 0.4) \cdot 10^{-4}$  &
 600 & SERP & $\pi^- p \to \eta n$, 1980~\cite{Dzhelyadin:1980kh}  \\
\hline
$\eta^\prime \to e^+e^-\gamma$ & $(4.69 \pm 0.31) \cdot 10^{-4}$ & 864 &
 BES3 & $e^+e^- \to J/\psi \to \eta'\gamma$, 2015~\cite{Ablikim:2015wnx} \\ 
\hline
$\eta^\prime \to \mu^+\mu^-\gamma$ & $(1.08 \pm 0.27) \cdot 10^{-4}$ & 33 & 
SERP & 33~$\pi^- p \to \eta' n$, 1980~\cite{Dzhelyadin:1980ki}  \\
\hline
\end{tabular}
\end{table}
\end{center}
\vspace*{-1mm}
Decays to two lepton pairs offer a possibility to study TFF in other
$q^2$ ranges. We illustrate the current situation with the branching
determinations in Table~\ref{tab3}, which for completeness also comprises
decays into one lepton pair while another is replaced with a
$\pi^+\pi^-$ one. The latter also provide important info on TFF
assuming vector dominance. 

\begin{center}
\begin{table}[t]
\caption{\label{tab3}Status of $P \to l^+l^-l^{'+}l^{'-}$ Studies}
\begin{tabular}{|l|c|c|c|l|}
\hline
Decay mode & ${\cal {B}}$ & Events & Group & Process \\
\hline
$\pi^0 \to e^+e^-e^+e^-$ & $(3.38 \pm 0.16) \cdot 10^{-5}$ & 30.5k  & 
KTEV & $K^0_L \to \pi^0\pi^0\pi^0$, 2008~\cite{Abouzaid:2008cd} \\
\hline
$\eta \to e^+e^-e^+e^-$ & $(2.4 \pm 0.2) \cdot 10^{-5}$  & 362 & 
KLOE & $e^+e^- \to \phi \to \eta\gamma$, 2011~\cite{KLOE2:2011aa} \\
\hline
$\eta \to    e^+e^-\mu^+\mu^-$ & $< 1.6 \cdot 10^{-4}$  & 90\%CL &
 WASA & $p d \to \eta~^3$He, 2008~\cite{Berlowski:2007aa}  \\
\hline
$\eta \to    \mu^+\mu^-\mu^+\mu^-$ & $< 3.6 \cdot 10^{-4}$  & 90\%CL &
 WASA &  $p d \to \eta~^3$He, 2008~\cite{Berlowski:2007aa}   \\
\hline
$\eta \to e^+e^-\pi^+\pi^-$ & $(2.68 \pm 0.12) \cdot 10^{-4}$  &
1555 &  KLOE & $e^+e^- \to \phi \to \eta \gamma$, 2009~\cite{Ambrosino:2008cp}  \\
\hline
$\eta \to    \mu^+\mu^-\pi^+\pi^-$ & $< 3.6 \cdot 10^{-4}$  & 90\%CL &
 WASA &  $p d \to \eta~^3$He, 2008~\cite{Berlowski:2007aa}   \\
\hline
$\eta' \to e^+e^-\pi^+\pi^-$ & $(2.11 \pm 0.18) \cdot 10^{-3}$  &
429 & BES3 & $e^+e^- \to J/\psi \to \eta' \gamma$, 
2013~\cite{Ablikim:2013wfg}  \\
\hline
$\eta' \to \mu^+\mu^-\pi^+\pi^-$ & $< 0.29  \cdot 10^{-4}$  & 90\%CL
& BES3 & $e^+e^- \to J/\psi \to \eta' \gamma$, 2013~\cite{Ablikim:2013wfg}  \\
\hline
\end{tabular}
\end{table}
\end{center}
\vspace*{-1mm}

Finally, in Table~\ref{tab4} we list the most precise measurements of the
branchings for pseudoscalar decays to $V\gamma$ and $Vl^+l^-$, which in
the Vector Dominance Model provide complementary information on TFF.
  
\begin{center}

\begin{table}[h]
\caption{\label{tab4}Status of $P \to V \gamma$ and $P \to V l^+l^-$ Studies}
\begin{tabular}{|l|c|c|c|c|}
\hline
Decay mode & ${\cal {B}}$ & Events & Group & Process \\
\hline
$\eta \to \pi^+\pi^-\gamma$ & $(4.22 \pm 0.08) \cdot 10^{-2}$  & 200k & KLOE 
& $e^+e^-  \to \phi \to \eta\gamma$, 2013~\cite{Babusci:2012ft1} \\
\hline
$\eta' \to \pi^+\pi^-\gamma$ & $(29.2 \pm 0.5) \cdot 10^{-2}$  & 200 & CLEO 
& $e^+e^-  \to J/\psi \to \eta'\gamma$, 2009~\cite{Pedlar:2009aa} \\
\hline
$\eta' \to \omega\gamma$ & $(2.55 \pm 0.16) \cdot 10^{-2}$  & 33.2k & 
BES3 & $e^+e^-  \to J/\psi \to \eta'\gamma$, 2015~\cite{Ablikim:2015eos} \\
\hline
$\eta' \to \omega e^+e^-$ & $(1.97 \pm 0.38) \cdot 10^{-4}$  & 66 & 
BES3 & $e^+e^-  \to J/\psi \to \eta'\gamma$, 2015~\cite{Ablikim:2015eos} \\
\hline
\end{tabular}
\end{table}
\end{center}

\newpage

\subsection{ Meson Transition Form Factors from Hadronic Processes}
\addtocontents{toc}{\hspace{2cm}{\sl A. Kupsc}\par}

\vspace{5mm}

A. Kupsc

\vspace{5mm}
\noindent
Department of Physics and Astronomy\\
 Uppsala University, Sweden
\vspace{5mm}

Transition form factors of neutral pseudoscalar mesons $P=\pi^0$,
$\eta$ or $\eta'$, describing $P\gamma^{(*)}\gamma^{(*)}$ vertex,
provide necessary input to calculate dominant part of the hadronic
light-by-light contribution to the anomalous magnetic moment of the
muon, $a_{\mu}$.  Ultimately the information about double off-shell
transition form factors is needed. Experimentally it would involve
challenging experiments like double tagged pseudoscalar meson
production $e^+e^-\to e^+e^- P$. An interesting alternative to extract
the information from much more common hadronic/radiative processes is
provided by dispersion relations. The processes where one of the
virtual photons is replaced by a hadronic system in $J^{PC}=1^{--}$ state 
could be used.

The simplest example is $\eta/\eta'\to\pi^+\pi^-\gamma$
process which is related to Dalitz decay  $\eta/\eta'\to e^+e^-\gamma$ 
and therefore to single off shell $\eta/\eta'$ transition
form factor \cite{Stollenwerk:2011zz,Hanhart:2013vba,Kubis:2015sga}. 
Experimental input is provided by studying invariant mass
distribution of dipion system in  $\eta/\eta'\to\pi^+\pi^-\gamma$ 
decays. In case of $\eta$ meson decays the distributions were measured
in high statistics experiments WASA-at-COSY \cite{Adlarson:2011xb} and KLOE
\cite{Babusci:2012ft}. The corresponding high statistics distributions
are expected soon from BESIII experiment. 

The dispersive approach is also being applied to determine $\pi^0$
transition form factor from $e^+e^-\to\pi^0\pi^+\pi^-$ and
$e^+e^-\to\pi^0\omega$ processes \cite{Hoferichter:2014vra}. It was 
also proposed to extend it to the $\eta$ form factor by considering
$e^+e^-\to\eta\pi^+\pi^-$ process \cite{Xiao:2015uva}.

\newpage

\subsection{On the precision of a data-driven estimate of 
hadronic light-by-light scattering in the muon $g-2$: \\ 
pseudoscalar-pole contribution}
\addtocontents{toc}{\hspace{2cm}{\sl A.~Nyffeler}\par}

\vspace{5mm}

A.~Nyffeler

\vspace{5mm}
\noindent
Institut f\"ur Kernphysik and PRISMA Cluster of Excellence, \\ 
Johannes Gutenberg-Universit\"at Mainz,   
D-55128 Mainz, Germany \\
\vspace{5mm}

For many years now there is a discrepancy of $3-4$ standard deviations
between the Standard Model prediction for the muon $g-2$ and the
experimental value~\cite{JN_09,g-2_reviews_exp}. This might be a sign
of New Physics, but there are also some doubts about the size and
uncertainties of the hadronic contributions which should be controlled
better~\cite{g-2_hadronic} in view of planned new $g-2$ experiments at
Fermilab and J-PARC~\cite{future_g-2_exp}. In particular hadronic
light-by-light scattering (HLbL) has so far only be estimated using
hadronic models, see Ref.~\cite{HLbL_review} for a recent
overview. Some attempts are ongoing to calculate HLbL using Lattice
QCD~\cite{HLbL_Lattice}. In this situation, recently a data-driven
dispersive approach to HLbL was proposed by two groups~\cite{HLbL_DR},
which relates the numerically dominant contributions from the
pseudoscalar-{\it poles} and the pion-loop with {\it on-shell}
intermediate pseudoscalars states to, in principle, measurable form
factors and cross-sections with off-shell photons: $\gamma^* \gamma^*
\to \pi^0, \eta, \eta^\prime$ and $\gamma^* \gamma^* \to \pi^+ \pi^-,
\pi^0 \pi^0$.

Within this dispersive framework the pseudoscalar-pole contribution
$a_\mu^{\HLbLP}$ with ${\rm P} = \pi^0, \eta, \eta^\prime$ was studied
recently in great detail in Ref.~\cite{Nyffeler_16} and the main
results were presented at this meeting. This contribution to HLbL can
be evaluated using a three-dimensional integral
representation~\cite{JN_09} with $a_\mu^{\HLbLP} = \left( \alpha / \pi
\right)^3 \left[ a_\mu^{\HLbLPone} + a_\mu^{\HLbLPtwo} \right]$, where
\bea 
a_\mu^{\HLbLPone} & = & \int_0^\infty \!\!\!dQ_1 \!\!\int_0^\infty
\!\!\!dQ_2 \!\!\int_{-1}^{1} \!\!d\tau \, w_1(Q_1,Q_2,\tau) \, 
\FFP(-Q_1^2, -(Q_1 + Q_2)^2) \, \FFP(-Q_2^2,0), \nonumber \\ 
a_\mu^{\HLbLPtwo} & = & \int_0^\infty \!\!\!dQ_1 \!\!\int_0^\infty
\!\!\!dQ_2 \!\!\int_{-1}^{1} \!\!d\tau \, w_2(Q_1,Q_2,\tau) \,  
\FFP(-Q_1^2, -Q_2^2) \, \FFP(-(Q_1+Q_2)^2,0), \nonumber 
\eea 
which involves the single-virtual $\FFP(-Q^2, 0)$ and double-virtual
$\FFP(-Q_1^2, -Q_2^2)$ pseu\-do\-sca\-lar-photon transition form
factor (TFF) for spacelike momenta. The integrations run over the
lengths of the two spacelike (Euclidean) four-momenta $Q_1$ and $Q_2$
and the angle $\theta$ between them $Q_1 \cdot Q_2 = Q_1 Q_2
\cos\theta$ with $\tau = \cos\theta$.

\begin{figure}[t!]
\centerline{\includegraphics[width=0.55\textwidth]{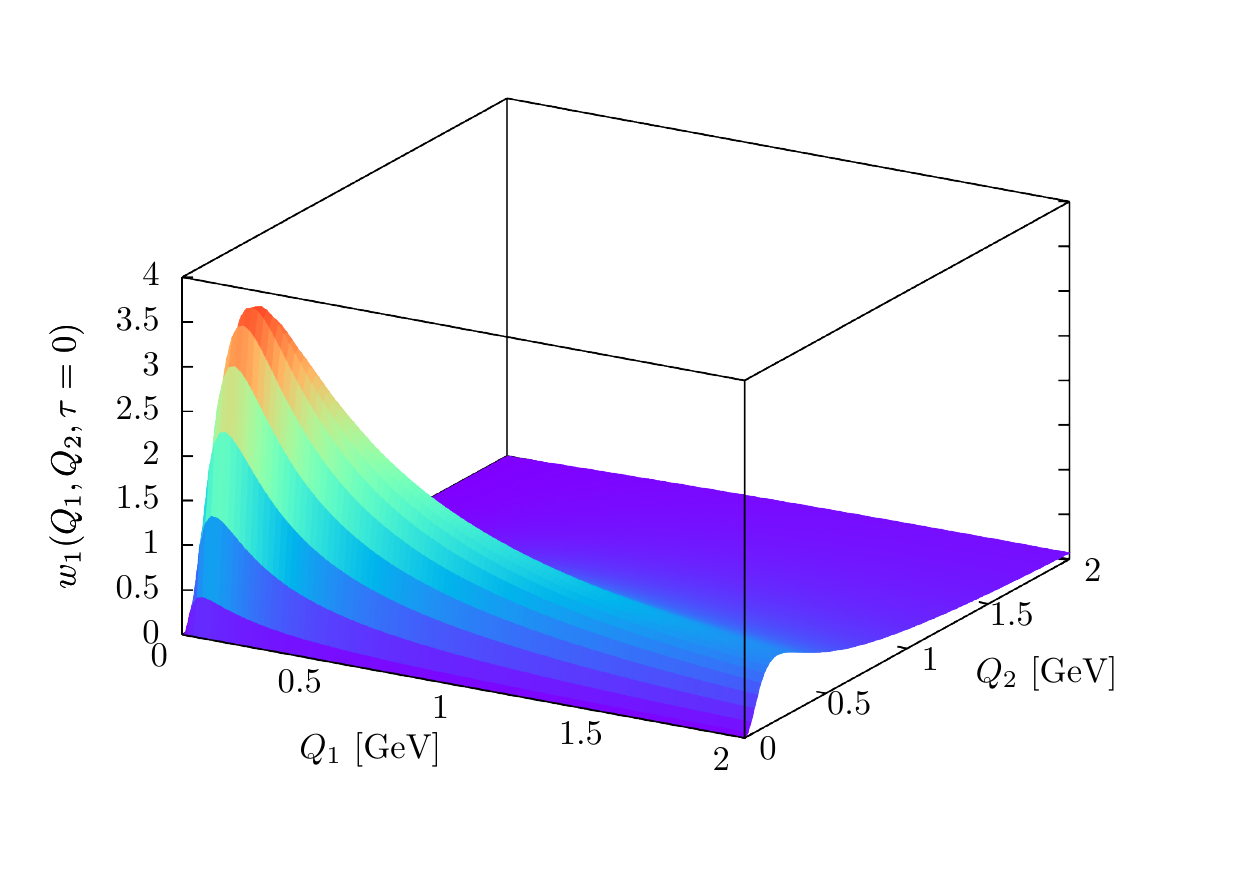}
\hspace{-10mm}
\includegraphics[width=0.55\textwidth]{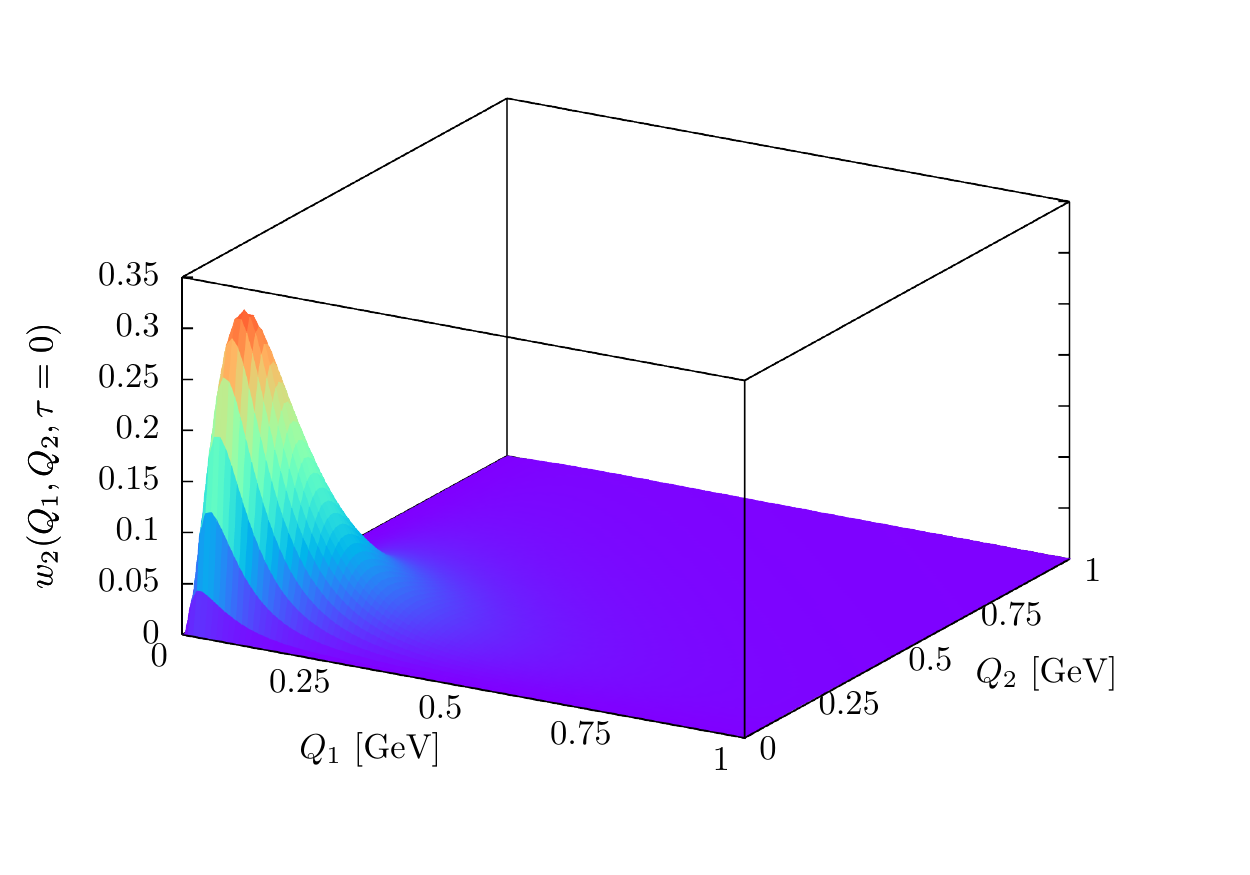}}
\caption{Model-independent weight functions $w_{1}(Q_1,Q_2,\tau)$
  (left) and $w_{2}(Q_1,Q_2,\tau)$ (right) for the pion as function of
  the Euclidean momenta $Q_1$ and $Q_2$ for $\theta = 90^\circ~(\tau =
  0)$.}
\label{Fig:w_i_pion}
\end{figure}

The three-dimensional integral representation separates the generic
kinematics, described by the model-independent dimensionless weight
functions $w_{1,2}$, which include the pseudoscalar propagator, from
the dependence on the single- and double-virtual TFF. For the pion,
the weight functions are concentrated below about 1~GeV, see
Fig.~\ref{Fig:w_i_pion}, i.e.\ the low-momentum region will be
dominant for the pion-pole contribution to HLbL. For $\eta$ and
$\eta^\prime$ the weight functions are spread out a bit to higher
momenta of around $1.5-2$~GeV. Note that for $\theta \leq
150^\circ~(\tau \geq -0.85)$ there is a ridge in $w_1(Q_1, Q_2, \tau)$
which leads to a divergent result for $a_\mu^{\HLbLPone}$ with
constant Wess-Zumino-Witten form factors.  To obtain finite results
for $a_\mu^{\HLbLPone}$, some damping from form factors is needed and
the analysis in Ref.~\cite{Nyffeler_16} is based on two simple models
for the TFF, vector meson dominance (VMD) and, for the pion, lowest
meson dominance plus an additional vector multiplet
(LMD+V)~\cite{LMDV}, which has a better matching with perturbative
QCD and the OPE at high momenta than the VMD
model. Ref.~\cite{Nyffeler_16} contains details which momentum bins in
the $(Q_1, Q_2)$-plane contribute how much to $a_\mu^{\HLbLP}$ when
integrated over all angles or how $a_\mu^{\HLbLP}$ changes when one
integrates up to some momentum cutoff. For $\pi^0$~[$\eta$,
$\eta^\prime$], the bulk of the contribution (around $85-95\%$, the
precise number is model-dependent) comes from the region below about
$1~[1.5]$~GeV.

Furthermore, we analyzed in Ref.~\cite{Nyffeler_16} how the precision
of current and future measurements of the single- and double-virtual
TFF in different momentum regions impacts the precision of such a
data-driven estimate of the pseudoscalar-pole contribution to
HLbL. For the single-virtual TFF of the pion $\FFa(-Q^2, 0)$ there are
only data available in the spacelike region at momenta above $0.7$~GeV
with rather large uncertainties in some momentum bins. Hopefully, the
situation will be improved soon by upcoming measurements by BESIII 
above $0.5$~GeV~\cite{BESIII_single_virtual}. For $\eta$ and
$\eta^\prime$ there is also information on the TFF available in the
timelike region from single Dalitz decays. On the other hand, there
are currently no measurements for the double-virtual TFF for none of
the pseudoscalars. Based on Monte Carlo simulations for a planned
first measurement of these double-virtual form factors at BESIII, with
bin-wise statistical uncertainties listed in Ref.~\cite{Nyffeler_16},
we find that combined with present and upcoming information on the
single-virtual form factors, a precision of
\bea 
\delta a_\mu^{\HLbLpi} / a_\mu^{\HLbLpi} & = & 14\% \quad
[11\%], \nonumber \\   
\delta a_\mu^{{\rm HLbL};\eta} / a_\mu^{{\rm HLbL};\eta} & = & 
23\%,  \nonumber \\   
\delta a_\mu^{{\rm HLbL};\eta^\prime} / a_\mu^{{\rm
      HLbL};\eta^\prime} & = & 15\%, \nonumber   
\eea 
seems feasible for the $\pi^0, \eta, \eta^\prime$-pole contributions
to HLbL. The result in bracket for the pion uses a dispersion
relation~\cite{DR_pion_TFF} for the single-virtual TFF below 1~GeV.

Compared to the range of estimates using various models in the
literature~\cite{JN_09,g-2_reviews_exp,HLbL_review,PdeRV_09},
$a_{\mu;{\rm models}}^{\HLbLpi} = (50 - 80) \times 10^{-11} = (65 \pm
15) \times 10^{-11}~(\pm 23\%)$ and $a_{\mu;{\rm models}}^{\HLbLP} =
(59 - 114) \times 10^{-11} = (87 \pm 27) \times 10^{-11}~(\pm 31\%)$,
this would definitely be some progress, as it would be largely based
on experimental input data only. More work is needed, however, to
reach a precision of $10\%$ for all three contributions which is
envisioned in the data-driven approach to HLbL. Further improvements
can be expected from more precise experimental data and from the use
of dispersion relations and Lattice QCD for the different form factors
themselves.

\section*{Acknowledgments} 

I am very grateful to Achim Denig, Christoph Redmer and Pascal Wasser
for providing me with results of Monte Carlo simulations for planned
transition form factor measurements at BESIII and to Martin
Hoferichter and Bastian Kubis for sharing information about the
precision of the dispersive approach to the pion TFF. This work was
supported by Deutsche Forschungsgemeinschaft (DFG) through the
Collaborative Research Center ``The Low-Energy Frontier of the
Standard Model'' (SFB 1044).

\newpage

\subsection{Deuteron Electromagnetic Structure in Holographic QCD}
\addtocontents{toc}{\hspace{2cm}{\sl V.E.~Lyubovitskij}\par}

\vspace{2mm}

V.E.~Lyubovitskij$^{1,2,3}$, 
T.~Gutsche$^1$, I.~Schmidt$^4$  

\vspace{2mm}

\noindent
$^1$ 
Institut f\"ur Theoretische Physik,
Universit\"at T\"ubingen, 
Kepler Center for Astro and Particle Physics,
Auf der Morgenstelle 14, D-72076 T\"ubingen, Germany\\
$^2$ 
Department of Physics, Tomsk State University,
634050 Tomsk, Russia\\
$^3$ 
Laboratory of Particle Physics,
Mathematical Physics Department, \\
Tomsk Polytechnic University,
Lenin Avenue 30, 634050 Tomsk, Russia \\
$^4$ 
Departamento de F\'\i sica y Centro Cient\'\i
fico Tecnol\'ogico de Valpara\'\i so (CCTVal), Universidad T\'ecnica
Federico Santa Mar\'\i a, Casilla 110-V, Valpara\'\i so, Chile

\vspace{4mm}

We apply a soft-wall AdS/QCD 
approach~\cite{AdSQCD} to a description of
deuteron form factors and structure functions.
It completes our previous study in~\cite{Gutsche:2015xva}. 
By appropriate choice of the two couplings in the effective action 
we are able to produce the form factors and structure functions in full consistency with
constraints derived in pQCD~\cite{Brodsky:1983vf}. 
The results are shown in Fig.~1. 
Our predictions for charge $r_C = 1.92$~fm 
and magnetic $r_M = 2.24$~fm radii 
compare well with the data: 
$r_C = 2.13 \pm 0.01$~fm and
$r_M = 1.90 \pm 0.14$~fm.

\begin{figure}[h]
\begin{center}
\vspace*{-.3cm}
\includegraphics[scale=.5]{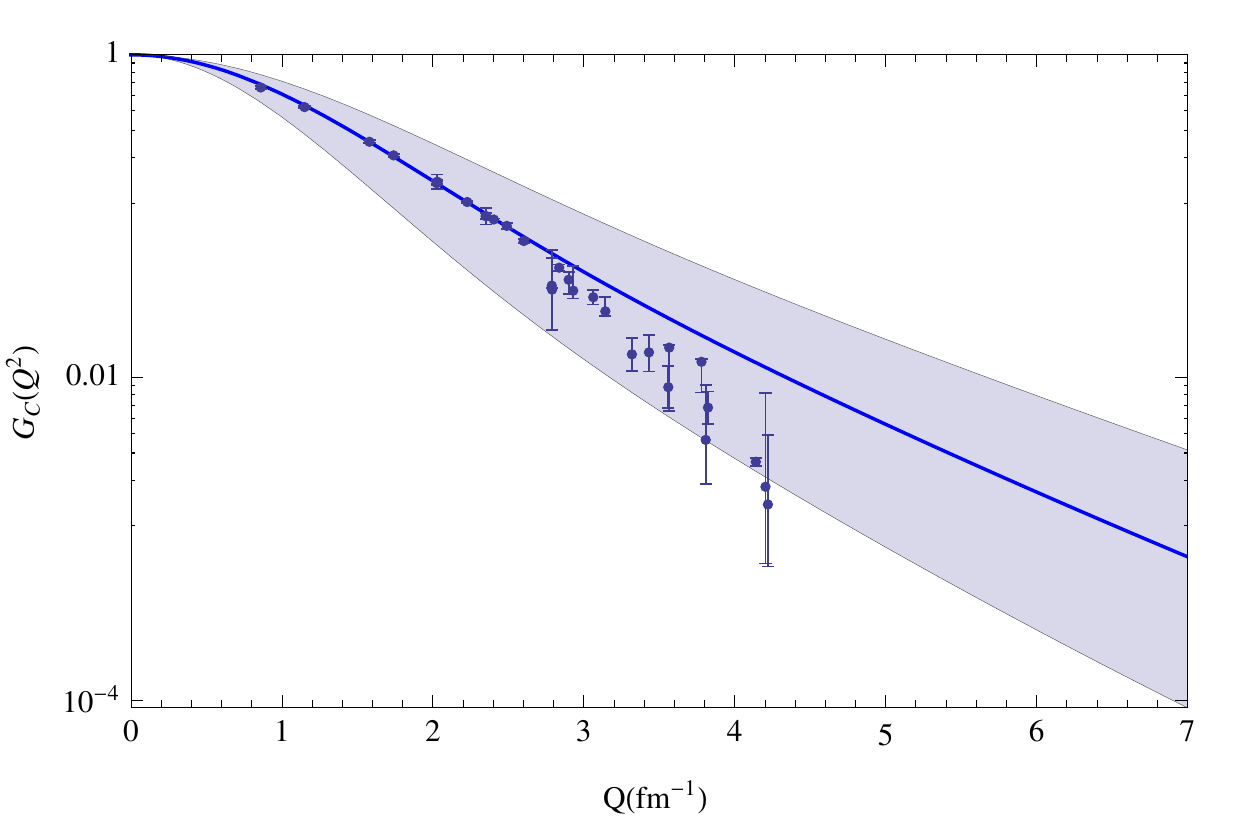}
\includegraphics[scale=.5]{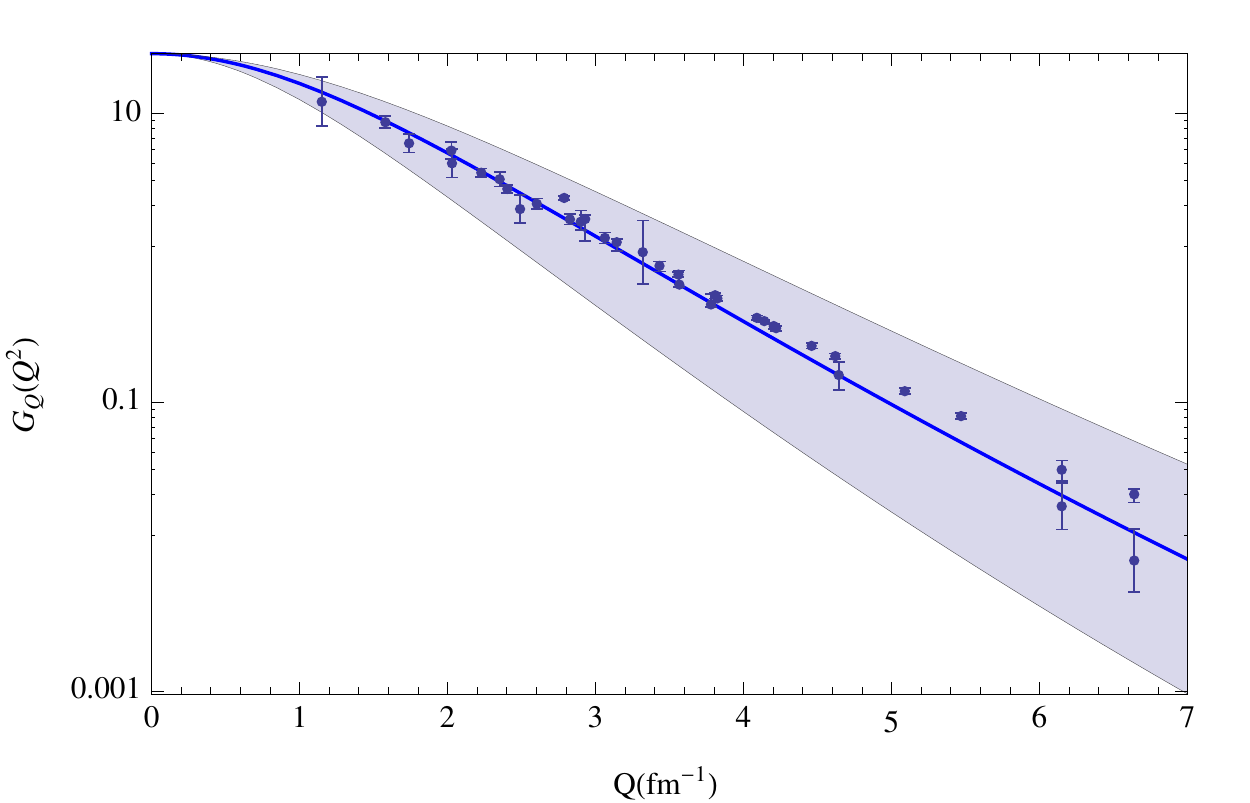}
\includegraphics[scale=.5]{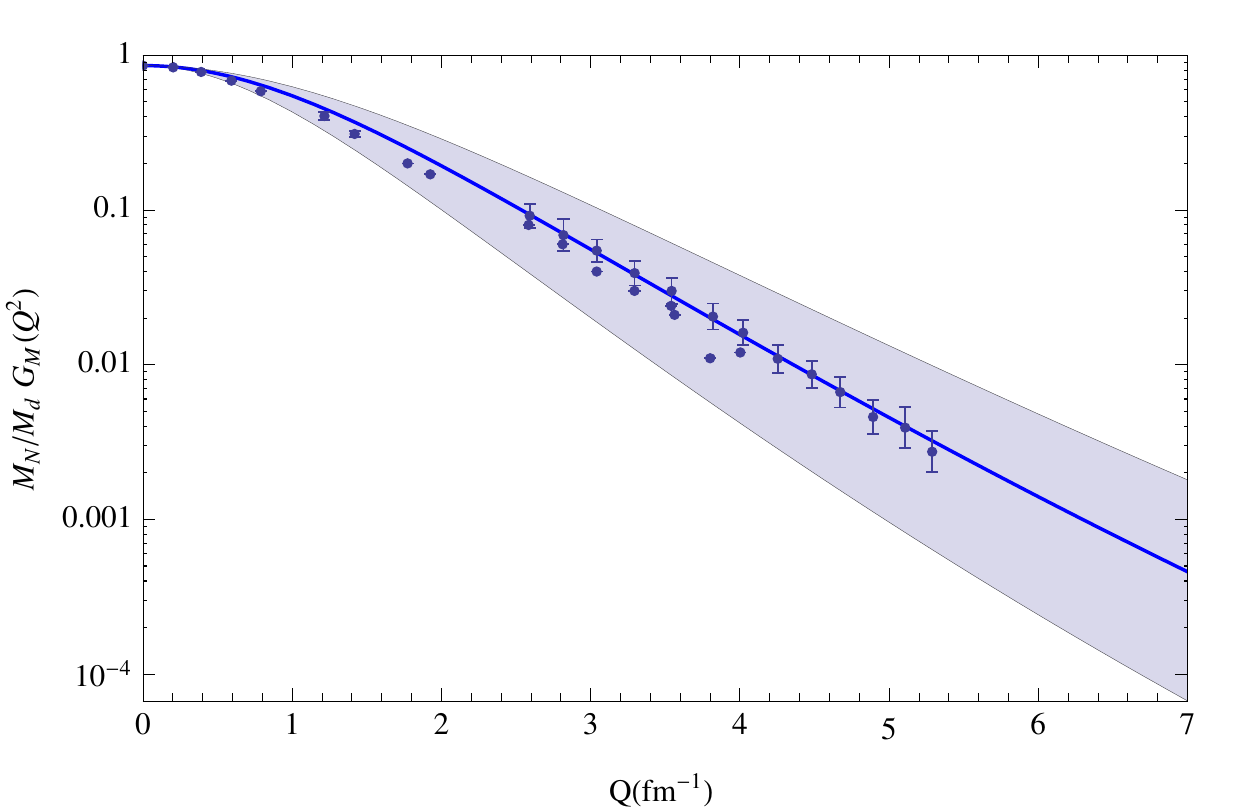}
\includegraphics[scale=.5]{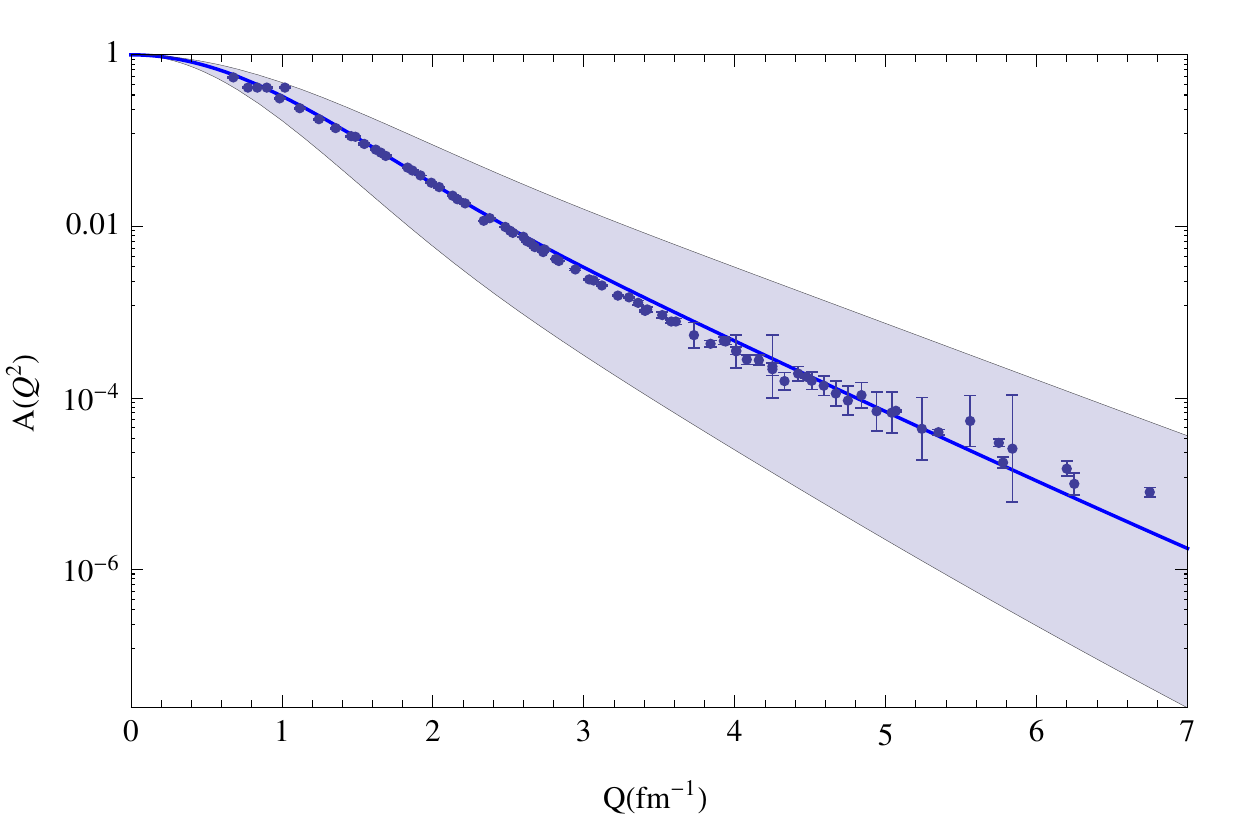}
\includegraphics[scale=.5]{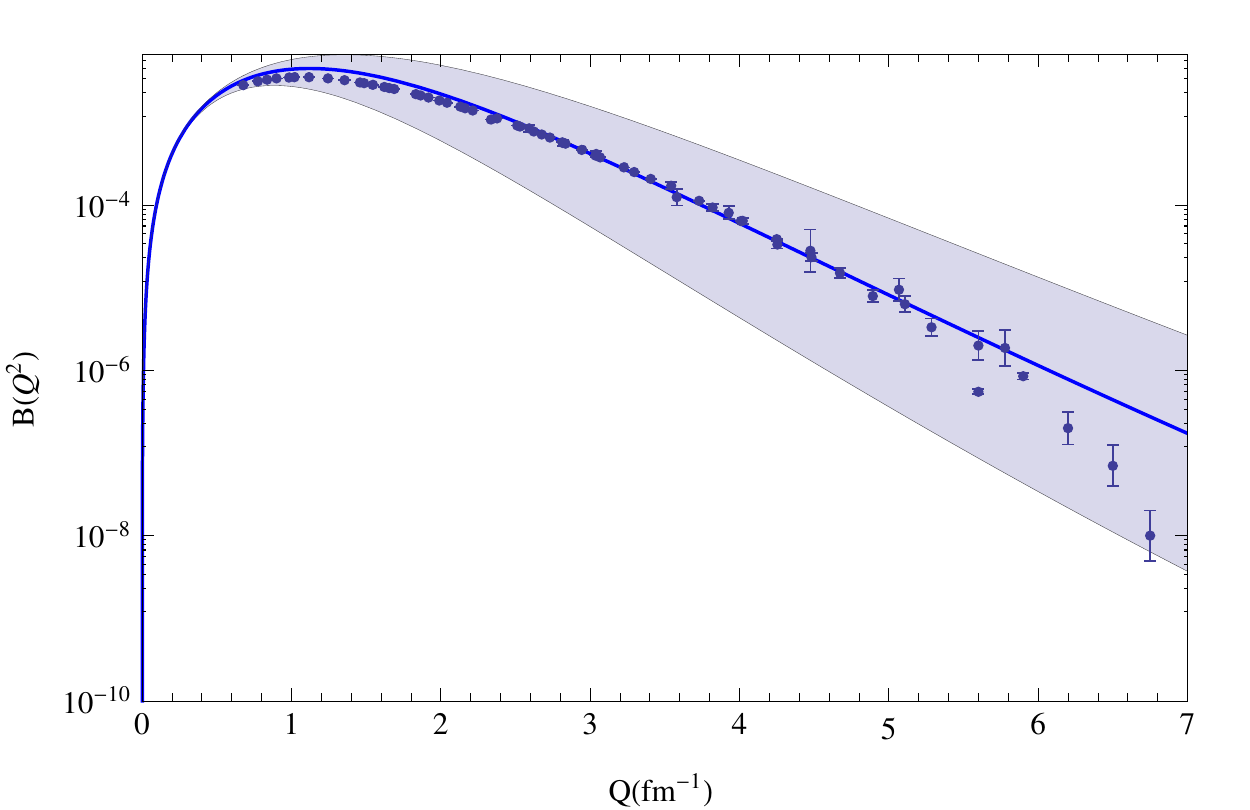}
\caption{Deuteron form factors and structure functions.} 
\end{center}
\end{figure}

This work was supported
by the BMBF under Project 05P2015 (FSP 202), 
by CONICYT Research Project No. 80140097,
by FONDECYT under Grants No. 1140390 and FB - 0821,
by TSU CIP and RF program ``Nauka'' (No. 0.1526.2015, 3854).


\newpage

\subsection{$V \to P \gamma^*$ transitions from rational approximants}
\addtocontents{toc}{\hspace{2cm}{\sl P.~Masjuan}\par}

\vspace{5mm}

P.~Masjuan and P.~Sanchez-Puertas

\vspace{5mm}
\noindent
PRISMA Cluster of Excellence, Institut f\"ur Kernphysik, Johannes Gutenberg-Universit\"at Mainz,
D-55099 Mainz, Germany
 \\
\vspace{5mm}

In this presentation we explore the idea that electromagnetic transitions between vector and pseudoscalar mesons, i.e., $V \to P \gamma^*$ transitions, can as well be seen as the decay of a pseudoscalar into two virtual photons in the time-like (TL) region for one of the photons converted to a vector meson, i.e, $P \to V \gamma^*$. From this point of view, one can exploit experimental data on the space-like (SL) region for the pseudoscalar transition form factor (TFF) and analytically continue to the TL to predict the energy behavior of the virtual photon giving rise to the $VP$ form factor (FF).

In recent works~\cite{Masjuan:2015lca} we explored the possibility to describe the double virtual TFF using as input information the low-energy parameters (LEP) of the single virtual TFF together with a mathematical tool to reconstruct from these LEPs the desired TFF~\cite{Masjuan:2008fv,Masjuan:2012wy,Escribano:2013kba,Escribano:2015nra,Escribano:2015vjz,Escribano:2015yup}.

This mathematical technic is an extension of Pad\'e approximants (PA)~\cite{Baker} to approximate functions of double virtuality and is called Canterbury approximants (CA)~\cite{Chisholm}. This technic is based on a well defined mathematical problem, the \textit{general rational Hermite interpolation problem}. This problem corresponds with the situation where a function should be approximated or reconstructed but information about it is scarce and spread over a set of points and derivatives. Our goal is to make a contact with this problem from our needs and provide a data-driven parameterization of the TFFs. The method proposed is, indeed, a \textit{method}~\cite{Baker}, not a model, simple~\cite{Masjuan:2008fv}, easy to understand, to apply and reproduce. It may contain approaches (improvable), but not assumptions (not improvable). It is systematic~\cite{Masjuan:2012wy,Escribano:2013kba,Escribano:2015nra,Escribano:2015yup,Masjuan:2007ay}, easy to update with new experimental data or limiting constraints~\cite{Escribano:2015nra,Escribano:2015yup} but also provides with a systematic error, a pure error from the method itself~\cite{Masjuan:2012wy,Escribano:2013kba,Escribano:2015nra,Escribano:2015yup,Masjuan:2007ay}. Finally, it is predictive~\cite{Escribano:2015nra,Escribano:2015yup,Aguar-Bartolome:2013vpw}.

The corpus of Pad\'e Theory~\cite{Baker} defines precisely the problem we have at hand and provides with a solution in the form of convergence theorems and tools. Parameterizations such as Vector Meson Dominance, Lowest Meson Dominance (and extensions), models from holographic QCD are already a certain kind of PAs (the so-called Pad\'e-Type approximants~\cite{Baker} where the poles of the Pad\'e are given in advanced). These parameterizations should take advantage of the theory of PAs if a robust calculation is to be claimed. For example, the uncertainties due to the truncation of the PA sequence which are never accounted for when using these parameterizations, do not need to be small to be neglected. Moreover, it is proven that Pad\'e-Type converge slower than PAs~\cite{Queralt:2010sv}.

The analytic continuation of PAs from SL to TL has been proven recently~\cite{Escribano:2015nra} to be valid only for TFF where the imaginary part in the TL is driven by vector states and rescatering of pions in their P-wave. Actually, the $V \to P \gamma^*$ is an excellent laboratory to test the performance of PAs in the TL for virtual photons. Not only that, but given the fact that the  $\omega$ is so narrow (its propagator is a meromorphic function), and the $\rho$ has a branch cut which imaginary part is positive defined, the PA's convergence theorems are at work~\cite{Baker,Masjuan:2007ay}.

The application of CAs to the $\omega \to \pi^0 \mu^+\mu^-$ case for which experimental data is available from the NA60 Collaboration~\cite{Arnaldi:2016pzu} is as follows. (Details of the calculation, comparison with other models, and prospects will be given elsewhere~\cite{inprep}.) To reproduce the experimental result, first we take the first virtual photon to be exactly at the $\omega$ mass so the vector meson seats on its mass shell. Then, to reproduce the invariant mass of the lepton pair we shall realize that the slope of the $\pi^0$ TFF measured in SL kinematics~\cite{pi0data} is not the same as the slope of the invariant mass of the lepton pair in the  $\omega \to P \mu\mu$ since the former is Bose symmetric and the latter is not due to isospin breaking. To account for this fact, split the slope $b_{\pi}$ of the SL $\pi^0$ TFF~\cite{Masjuan:2012wy} extracted from~\cite{pi0data} into a component coming from each allowed isospin, which is the $\omega$ component separated from the $\rho$ component. Using a quark model~\cite{Landsberg:1986fd} we conclude 
\begin{equation}
\frac{b_{\pi}}{m_{\pi}^2} = \frac{ b_{V}}{M_{\rho}^2}\left( \frac12 + \frac{ M_{\rho}^2}{2 M_{\omega}^2 b_{V}}\right)\, .
\end{equation}

\begin{wrapfigure}{l}{10cm}
\includegraphics[width=0.6\textwidth]{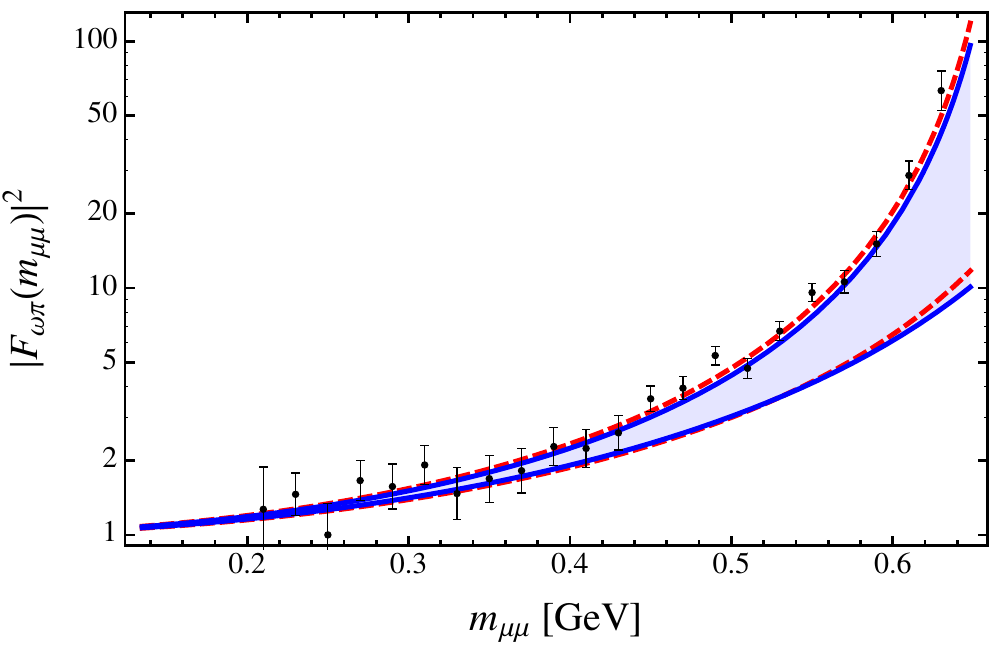}
\caption{$|F_{\omega\pi^0}(m_{\ell\bar{\ell}})|^2$ in terms of the invariant mass $m_{\ell\bar{\ell}}$ predicted with $F^0_1(q_2)$ and $F^1_2(q_2)$ (red and blue bands resp.). Experimental data from NA60 Collaboration~\cite{Arnaldi:2016pzu}.}
\label{omegaTFF}
\end{wrapfigure}

This implies that the slope corresponding to the lepton-pair invariant mass which has only isospin $I=1$ contribution should have a slope $b_V=1.16(15)$. This slope is equivalent as having an effective vector mass with $I=1$ of around $720(50)$MeV. This small mass explains the large raising of the experimental data at high energies and also why including exclusively the $\rho$ meson is not enough to account for the whole FF. With the same argument, one can calculate the corresponding curvature $c_V=1.20(26)\cdot 10^{-3}$.

To calculate the normalized-to-one $F_{\omega\pi}(m_{\mu\mu})$ we use the first and second elements of the CA sequence with the correct high-energy behavior which read
\begin{eqnarray}\label{Chisholm}
F^0_1(q_2)= \lim_{q^2_1 \to m_{\omega}^2}(q_1^2-m_{\omega}^2)C^0_1(q_1^2,q_2^2)=\frac{1}{1-b_{\rho}q_2^2}\, ,\\
F^1_2(q_2)= \lim_{q^2_1 \to m_{\omega}^2}(q_1^2-m_{\omega}^2)C^1_2(q_1^2,q_2^2)=\frac{1- a q_2^2}{1-b q_2^2+ c q_2^4}\, ,
\end{eqnarray}
\noindent
with $a,b,c$ matched to $b_{\rho},c_{\rho}$  as described before and to $2 F_{\pi}$ after imposing the correct high-energy behavior. The corresponding results are reported in Fig.~\ref{omegaTFF}. The band shown comes from the error of the LEPs together with an error estimation on the quark model used of a $30\%$ which after combining in quadrature results negligible. 

We expect the results here discussed to make a positive impact in the evaluation of the $\pi^0$ contribution to the hadronic light-by-light piece of the $(g-2)_{\mu}$ as suggested in~\cite{Masjuan:2014rea}.


\newpage

\section{Documentary: Bruno Touschek with AdA in Orsay}

\subsection{Bruno Touscheck }
\addtocontents{toc}{\hspace{2cm}{\sl G.~Pancheri}\par}

\vspace{0mm}

G.~Pancheri$^1$ and L.~Bonolis$^{2}$

\vspace{2mm}
\noindent
$^1$~INFN Frascati  National Laboratories, Via E. Fermi 40, 00044 Frascati \\
$^{2}$~Max Planck Institute for the History of Science, Berlin, Germany
 \\

At this workshop, we have presented a documentary entitled {\it Bruno Touschek with AdA in Orsay}, which was prepared in 2013 to commemorate the first observation  of electron-positron collisions in a laboratory, obtained with the storage ring AdA,  Anello di Accumulazione. AdA had been built in Italy, at the  Frascati  National Laboratories, where it went into operation  in February 1961. In July 1962, because of limitations arising from   the injection system,  AdA  was   transported to Orsay. At  LAL, le Laboratoire de l'Acc\'el\'erateur Lin\'eaire, a team of Italian and French scientists, technicians and engineers obtained in 1964 the  first proof ever of electron-positron collisions \cite{amaldi,bernardini,bonolis,bonolispancheri,grecopancheri,bernardinipancheripellegrini}.

 \begin{figure}[h]
\centering
\resizebox{0.8\columnwidth}{!}{
\includegraphics{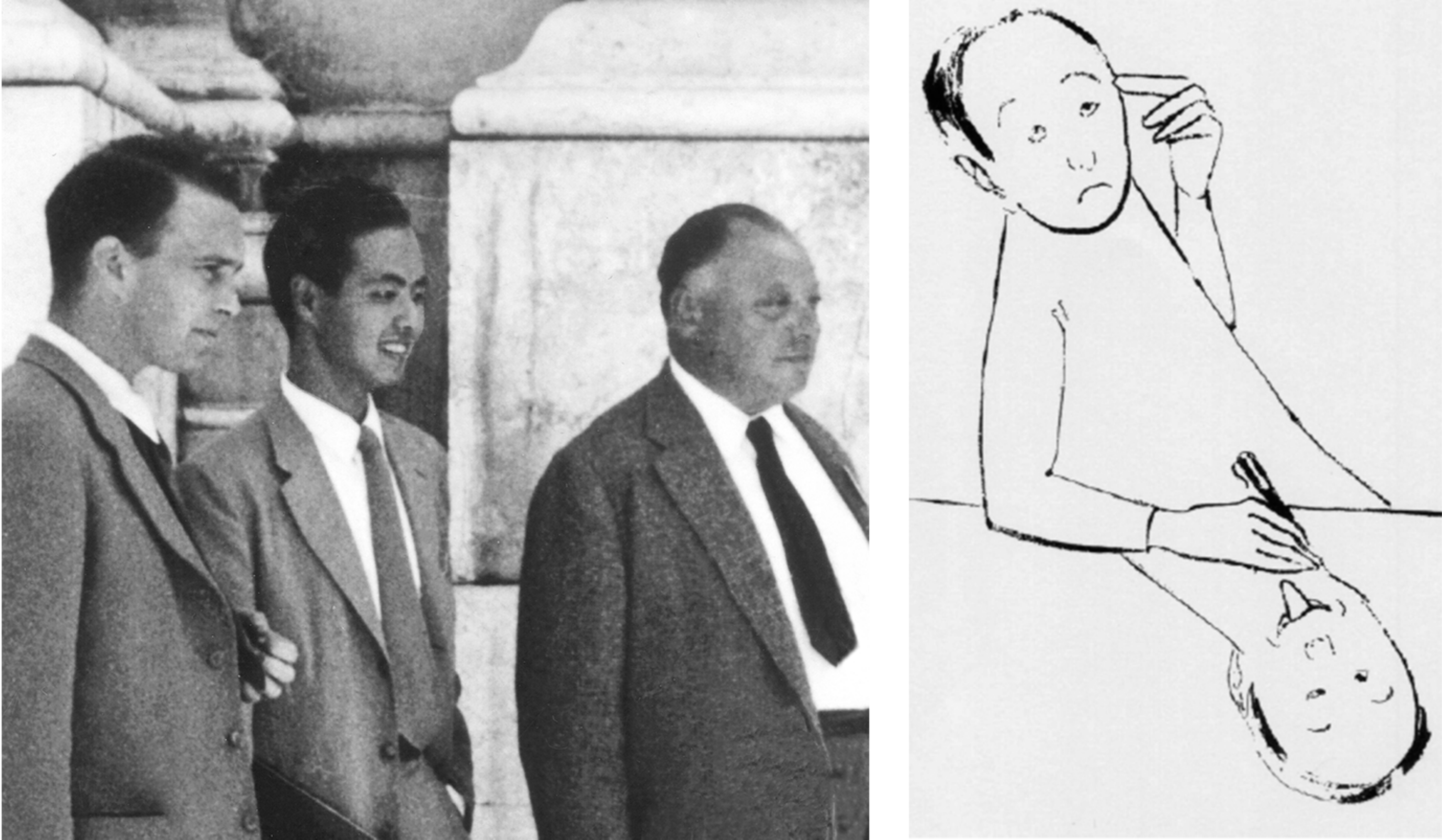}}
\caption{Bruno Touschek, T. D. Lee and Wolfgang Pauli in September 1957. On the right, Touschek's humorous drawing about T. D. Lee and Parity violation.}
\label{fig:1}

\end{figure}

\begin{figure}[h]
\centering
\resizebox{0.85\columnwidth}{!}{
\includegraphics{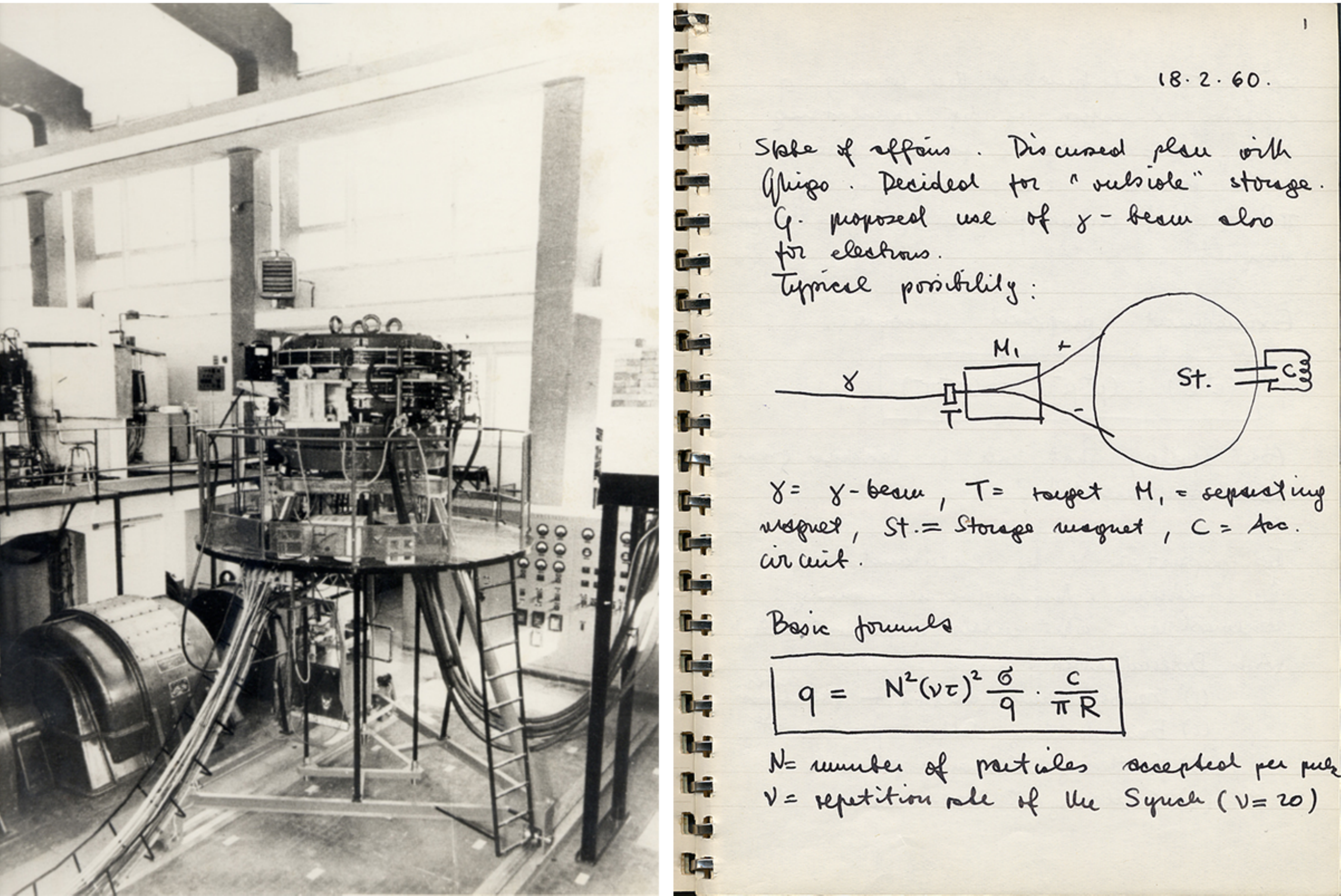}}
\caption{AdA in the hall hosting the electron-synchrotron in Frascati Laboratories. On the right, the first page of Touschek's notebook, starting on February 18, 1960.}
\label{fig:2}
\end{figure}

In Touschek, theoretical thought applied to experimental physics led to one of the  great ideas  of  accelerator physics, an {\it unthinkable idea}  which was  to shape  particle physics in the second half of the 20th century.  During the years in Rome, where he arrived in 1952, Touschek honed his theoretical skills, interacting with international visitors and Italian scientists.  In Fig. \ref{fig:1}, on the left, we show Touschek together with T.D. Lee and Wolfgang Pauli at the ``International Conference on mesons and recently discovered particles'' held in Padua and Venice in September 1957. In the  right panel, we show a drawing of TD Lee, illustrating in a humorous way the meaning of Parity violation, one of the hot issues of the conference. A few years later, Touschek would propose the construction of AdA and state his unshakeable faith in the CPT theorem. Carlo  Rubbia \cite{grecopancheri}, remembers meeting him in the dark corridors of the Guglielmo Marconi  Institute in Rome, saying with a loud voice  ``the positron and the electron must  collide because of the  CPT theorem!''. Touschek's theoretical genius and his training and understanding of the physics of particle accelerators, a very exceptional expertise at the times, acquired while working with   Rolf Wider\o e during WWII, is the combination of talents which led to the proposal of AdA, whose  picture next to the Frascati synchrotron is shown in the left hand panel of  Fig.  \ref{fig:2}.  In the right panel,  it is shown  the first page of the notebook where Touschek wrote down, for the first time on February 18th, 1960,  his ideas as to how and why to build an electron-positron accelerator.

\newpage

\section{Summaries of talks: Radiative corrections, form factors and MonteCarlo generators}

\subsection{Preliminary results on pion form factor at CMD-3}
\addtocontents{toc}{\hspace{2cm}{\sl F.~Ignatov}\par}

\vspace{2mm}
F.~Ignatov {\it for the CMD-3 Collaboration}

\vspace{2mm}
\noindent
Budker Institute of Nuclear Physics, Novosibirsk, Russia \\
\vspace{2mm}

The total $e^+e^- \to  hadrons$ cross section (or $R(s)$) is important for 
calculation of various physical quantities, e.g.
$\alpha_{QED}(M_Z)$ used in precise tests of EW physics
and better precision of this value is required in case of 
ILC. 
Also $R(s)$ is essential for the interpretation of precise measurements of
the anomalous magnetic moment of the muon $a_\mu = (g-2)/2$~\cite{Hagiwara:2011af}. The comparison of
this experimental value to the theoretical prediction provides a
powerful test of the Standard Model.

The dominant contribution to production of hadrons  in the energy range 
$\sqrt{s}<1$~GeV comes from the 
$e^+e^-\to\pi^+\pi^-$ mode. This channel gives the main contribution
to the hadronic term and overall theoretical precision of $a_\mu$. In
the light of new g-2 experiments at FNAL and J-PARC, which plan to reduce
an error by a factor of 4, it is very desirable to improve systematic
precision of the $\pi^+\pi^-$ cross section by at least a factor of two.

The CMD-3~\cite{Aulchenko:2001je,Khazin:2008zz}  detector has been successfully collecting data at the
electron-positron collider VEPP-2000~\cite{Shatunov:2000zc,Berkaev:2012qe} at Budker Institute of
Nuclear Physics since December 2010.
The first scan below 1 GeV for a $\pi^+\pi^-$ measurement was performed in 2013.
The collected  data sample corresponds to about 18 pb$^{-1}$ of integrated 
luminosity. 
The collected data sample is similar or larger than in
the previous experiments.

The geometry of $\pi^+\pi^-$ process allows a clean selection of $e^+e^-,
\mu^+\mu^-, \pi^+\pi^-$ collinear events by using the following criteria:
two collinear well reconstructed charged tracks are detected, these
tracks are close to the interaction point, fiducial volume of event is
inside a good region of the drift chamber. These final states can be separated using either the information about
energy deposition in the calorimeter or that about particle momenta in
the drift chamber.  At low energies
momentum resolution of the drift chamber is sufficient to separate different
types of particles. 

The difference between pion and electron momenta exceeds three standard
deviations for c.m. energies up to $2*E_{beam}\lesssim900\,\rm MeV$. 
At $\rho$ meson and higher energies the peak of electron shower in the calorimeter is
well distinguished from the peak of minimal ionization particles. 
The separation using energy deposition works best at higher
energies and becomes less robust at lower energies.
Determination of the number of different particles is done by
minimization of the binned likelihood function, where two dimensional PDF
functions are constructed in different ways for each type of information.
The two methods overlap in the wide energy range and provide a cross-check of 
each other, allowing to reach a systematic error of event separation at the level of 0.2\%.

The polar angle distribution can be used as additional input for
events separation by likelihood minimization.
A first attempt was made to include this information to PDFs constructed
with momentum distributions. It gives only small
statistical improvement, about 5\%, for $|F_{\pi}|^2 $ below $E_{beam}\lesssim420\,\rm
MeV$, but it can provides additional cross checks for systematics studies
of the event separation and of fiducial volume determination.

Another important source of systematics is a theoretical precision of
radiative corrections~\cite{Actis:2010gg}. Additional studies 
like crosschecks of different calculation approaches and further
proof from comparison with experimental data are necessary in this  
field. Comparison between the MCGPJ~\cite{Arbuzov:2005pt} and
BabaYaga@NLO~\cite{Balossini:2006wc} generators was performed. 
The integrated cross-section for applied cuts is well consistent at the 
level better than 0.1\%  between
two tools, but strong
difference in the $P^+ \times P^-$ momentum distributions was observed.
Also it was observed some discrepancy between experimental data and fitted 
functions when using event separation by momentum information, where
the initial input comes from the MCGPJ generator, while BabaYaga@NLO 
describes the data better.  While this discrepancy mostly  doesn't affect analysis by
energy deposition, it becomes crucial if momentum distribution information is used.
The current version of MCGPJ includes contributions from one hard
photon at large angle and any number of photons emitted exactly along initial and final
charged particles, while tails of the $P^+ \times P^-$ momentum
distributions with used selection criteria are determined mainly by contribution of two hard photons 
at large angles.
The MCGPJ generator improvements are now underway.
Addition  of the  angular distribution
for photon jets gives better consistency between predicted
distributions by MCGPJ and experimental data.
We expect that the overall uncertainty from MC tools can be reduced to
0.1\%. As seen from influence from two photon contributions to momentum
spectras, to reach precision $\lesssim 0.1\%$, 
it becomes very desirable to have exact $e^+e^- \to e^+e^- \gamma\gamma$ contribution inside generators, 

The final goal of the CMD-3 experiment is to reduce a overall systematic
uncertainty on the pion form factor measurement up to 0.35\%.

\newpage

\subsection{$(g-2)_{\mu}$: recent improvements and outlook}
\addtocontents{toc}{\hspace{2cm}{\sl A.~Keshavarzi}\par}

\vspace{5mm}

A.~Keshavarzi, T.~Teubner

\vspace{5mm}

\noindent
Department of Mathematical Sciences, University of Liverpool, Liverpool L69 3BX, U.K.\\
\vspace{0mm}

Since our last determination of the Standard Model (SM) value of the anomalous magnetic moment of the muon $a_{\mu}^{\mathrm{SM}}$ (HLMNT \cite{Hagiwara:2011af}), the standing $3-4\sigma$ discrepancy between $a_{\mu}^{\mathrm{SM}}$ and the experimental value $a_{\mu}^{\mathrm{exp}}$ \cite{Bennett:2006fi} has been further consolidated. With new experimental efforts at Fermilab \cite{Grange:2015fou} and J-PARC \cite{Iinuma:2011zz} aiming to reduce the uncertainty of $a_{\mu}^{\mathrm{exp}}$ by up to a factor of four, the uncertainty of the SM prediction must be improved. Should the existing discrepancy endure, it will be a strong signal of the existence of physics beyond the SM.

Of all the SM contributions, the leading order hadronic vacuum polarisation (HVP) contributions provide the largest uncertainty. The prospect of substantially improving the precision of $a_{\mu}^{\mathrm{HVP}}$ is a highly non-trivial task, with hadronic contributions from low energies excluding the use of perturbation theory and therefore largely depending on the precision of low energy hadronic cross section data. These experimental data are used as input into a dispersion integral (with a well known kernel function) which yields $a_{\mu}^{\mathrm{HVP}}$. However, the treatment and combination of the data plays a large role in the determination of the uncertainty of $a_{\mu}^{\mathrm{HVP}}$. The need to have a statistically valid and reliable data combination procedure is clearly evident.

The data combination procedure utilised in \cite{Hagiwara:2011af} combined an adaptive clustering algorithm to re-bin data points into clusters with a non-linear $\chi^2$ minimisation. Recent analyses of these highlighted the potential for systematic bias through the fitting of normalisation uncertainties (see, for example \cite{Ball:2009qv,Benayoun:2015gxa}). This, combined with the need to better include full experimental covariance matrices (such as those provided by the BaBar Collaboration \cite{Aubert:2009ad}), therefore necessitates the development of an improved data combination method.

We define, for each individual hadronic channel, a linear $\chi^2$ function depending on the fitted cluster centres, $R_m$,
\begin{equation} \label{NewChi^2}
\chi^2 (R_m) =\sum^{N_{\mathrm{tot}}}_{i=1}\sum^{N_{\mathrm{tot}}}_{j=1}\big(R_{i}^{(m)} - R_{m}\big) {\bf C}^{-1}\big(i^{(m)},j^{(n)}\big)\big(R_{j}^{(n)} - R_{n}\big) \ .
\end{equation}
Here, $N_{\mathrm{tot}}$ is the total number of data points, $R_{i}^{(m)}$ is the cross section value of the data point $i$ contributing to the cluster $m$, $R_{m}$ is the fitted cross section value of the cluster $m$ and ${\bf C}^{-1}\big(i^{(m)},j^{(n)}\big)$ is the inverse covariance matrix. Initially, the covariance matrix ${\mathrm {\bf C}}\big(i^{(m)},j^{(n)}\big)$ is fixed with the weighted averages of the cluster centres, $R_m^{0}$, such that
\begin{equation} \label{Ck}
{\bf C}\big(i^{(m)},j^{(n)}\big) = {\mathrm C}^{\mathrm{stat}}\big(i^{(m)},j^{(n)}\big) + {\mathrm C}^{\mathrm{sys}\%}\big(i^{(m)},j^{(n)}\big) R_{m}^{0}R_{n}^{0} \ ,
\end{equation}
where ${\mathrm C}^{\mathrm{sys}\%}\big(i^{(m)},j^{(n)}\big)$ is the matrix of percentage systematic uncertainties of each element. The $\chi^2$ minimisation procedure is then iterated, with each iteration's covariance matrix fixed with the previous iteration's fitted cluster centre values, until the routine converges.

Through toy models, the improved minimisation routine can be shown to be free from bias. It can also been shown that although the non-linear minimisation routine in \cite{Hagiwara:2011af} had the potential to exhibit systematic bias, previous results in \cite{Hagiwara:2011af,Hagiwara:2003da} were predominantly unaffected and are therefore still considered reliable. Overall, the new linear minimisation, together with a new clustering algorithm, results in a slightly lower mean value with a reduced uncertainty (and $\chi^2_{\mathrm{min}}/\mathrm{d.o.f.}$) across all hadronic channels.

After the inclusion of new data, we predict a much improved determination of $a_{\mu}^{\mathrm{HVP}}$. With the two-pion channel contributing over 70\% to $a_{\mu}^{\mathrm{HVP}}$ and its error, the inclusion of new data in this channel must be conducted with great scrutiny and care. Including new data from KLOE \cite{Babusci:2012rp}, the combination of the individual KLOE measurements through a combined covariance matrix \cite{DeLeo:2015yja} and a new measurement from BESIII \cite{Ablikim:2015orh}, we observe a reduction in the uncertainty of the two pion contribution of approximately one third. This, and the improvements in other hadronic channels, yields an overall uncertainty of $a_{\mu}^{\mathrm{HVP}}$ of less than half a percent. With new data in many channels expected to be made available in the near future, the determination of $a_{\mu}^{\mathrm{HVP}}$, and therefore $a_{\mu}^{\mathrm{SM}}$, should improve significantly in time for the new experimental measurements of $a_{\mu}$.

\newpage

\subsection{Measurement of the running of the fine structure constant below 1 GeV with the KLOE Detector}
\addtocontents{toc}{\hspace{2cm}{\sl G.~Venanzoni }\par}

\vspace{5mm}

G.~Venanzoni (on behalf of KLOE-2 Collaboration)

\vspace{5mm}

\noindent
 Laboratori Nazionali di Frascati dell'INFN, 00044 Frascati, Italy \\

\vspace{5mm}
Precision tests of the Standard Model (SM) require an appropriate inclusion of higher order effects and the  very precise knowledge of input parameters~\cite{Fred_2003}. One of the basic input parameters is the fine-structure constant $\alpha$, determined from the anomalous magnetic moment of the electron 
with the impressive accuracy of 0.37 parts per billion (ppb)~\cite{Aoyama}. 
However, physics at non-zero momentum transfer requires an effective electromagnetic coupling $\alpha(s)$. 
The shift of the fine-structure constant from
the Thomson limit to high energy involves low energy 
non-perturbative hadronic effects which spoil this precision. 

The vacuum polarization (VP) effects can be absorbed 
in a redefinition of the fine-structure constant, making it $q^2$ dependent:
\begin{equation}
\alpha(q^2) = {{\alpha(0)} \over {1-\Delta \alpha(s)}}
\label{running}
\end{equation}

The shift $\Delta\alpha$(s) in terms of the the vacuum polarization function $\Pi'_\gamma$(s) is given by:
\begin{equation}
\Delta \alpha(s) = -4\pi\alpha\,Re[\Pi_\gamma^{'}(s)-\Pi_\gamma^{'}(0)].
\label{delta_alpha}
\end{equation}
and it is the sum of the lepton ($e$,$\mu$,$\tau$) contributions, the 
contribution from the 5 light quark flavours (u,d,s,c,b) , and the contribution of the top quark (which can be neglected): $\Delta\alpha$(s)=$\Delta\alpha_{lep}(s)+
\Delta\alpha_{had}^{(5)}(s)+\Delta\alpha_{top}$.\\
The leptonic contributions can be calculated with very high precision in QED by the perturbation theory \cite{passera}.
However, due to the non-perturbative behavior of the strong
interaction at low energies, perturbative QCD only allows
us to calculate the high energy tail of the hadronic (quark)
contributions. In the lower energy region the hadronic contribution can be evaluated through a dispersion integral over the measured $e^+e^- \rightarrow$ hadrons cross-section:

\begin{equation}
\Delta\alpha_{had}(s)=-(\frac{\alpha s}{3\pi})Re\int_{ m_\pi^2}^\infty ds'\, \frac{R(s')}{s'(s'-s-i\epsilon)}
\end{equation}
where $R(s)$ is referred to the cross section ratio $R(s)=\frac{\sigma_{tot}(e^+e^-\rightarrow\gamma*\rightarrow hadrons)}{\sigma_{tot}(e^+e^-\rightarrow\gamma*\rightarrow \mu^+\mu^-)}$ .

In this approach the dominant uncertainty in the evaluation of $\Delta\alpha$ is given by 
the experimental data accuracy. 
Equations (\ref{running}) and (\ref{delta_alpha}) are the usual definition of the running
effective QED coupling and have the advantage that one obtains a real coupling.\\
The imaginary part of the VP function $\Pi'_\gamma$ is completely neglected, which is normally a good approximation as the contributions from the imaginary part are suppressed.
However, this approximation is not sufficient in the presence of resonances like the $\rho$ meson,
where the accuracy of the cross section measurements reaches the order of (or even less than) 1\%, and the imaginary part should be taken into account.\\
The KLOE-2 collaboration performed a measurement of the running of 
the fine structure constant $\alpha$
in the time-like region 0.6$<\sqrt{s}<$0.975 GeV~\cite{kloe2}. The strength of the coupling constant is measured differentially as a function of the momentum transfer of the
exchanged photon $\sqrt{s}=M_{\mu\mu}$ where $M_{\mu\mu}$ is the  $\mu^+\mu^-$ invariant mass.
The value of $\alpha(s)$ is extracted by the ratio of the  differential cross section 
 for the process $e^+e^- \rightarrow \mu^+\mu^- \gamma(\gamma)$ with the photon emitted in the Initial State (ISR) to the corresponding one from Monte Carlo (MC) simulation 
with the coupling set to the constant value $\alpha (s)$ $\equiv \alpha(0)$:
\begin{equation}
\vert \frac{\alpha(s)}{\alpha(0)}\vert^2= \frac{d\sigma_{data} (e^+e^- \rightarrow \mu^+\mu^- \gamma(\gamma))\vert_{ISR}/d\sqrt{s}}{d\sigma^{0}_{MC}(e^+e^- \rightarrow \mu^+\mu^- \gamma(\gamma))\vert_{ISR}/d\sqrt{s}}
\label{our_method}
\end{equation}

To obtain the ISR cross section, the observed cross section must be corrected for events with one or more photons in the final state (FSR). This has been done by using PHOKHARA MC event generator,  which includes  next-to-leading-order ISR and FSR contributions~\cite{PHOKHARA}.

We used events where the photon is emitted at small angles, which results in a large enhancement of the ISR with respect to the FSR contribution. 

From the measurement of the effective coupling constant and the KLOE dipion cross section \cite{Venanzo}, we extracted for the first time in a single experiment
the real and imaginary part of $\Delta\alpha$.

The analysis has been performed by using 
the data collected in 2004/05 by using the KLOE detector at DA$\Phi$NE, the $e^+e^-$ collider running at the $\phi$ meson mass, with a total integrated luminosity of 1.7 fb$^{-1}$.

\newpage

\subsection{Nuclean Form Factors: recent findings}
\addtocontents{toc}{\hspace{2cm}{\sl E.~Tomasi-Gustafsson }\par}

\vspace{5mm}

E.~Tomasi-Gustafsson$^1$, A.~Bianconi$^2$ and S.~Pacetti$^3$ 

\vspace{5mm} 
\noindent
$^1$~CEA,IRFU,SPhN, Saclay, 91191 Gif-sur-Yvette Cedex, France \\
$^2$~Dipartimento di Ingegneria dell'Informazione, 
Universit\`a di Brescia \\ and 
Istituto Nazionale di Fisica Nucleare, via A. Bassi 6, 
27100 Pavia, Italy 
 \\
$^3$~Dipartimento di Fisica e Geologia, and INFN Sezione di Perugia, 06123 Perugia, Italy
\vspace{5mm} 

The purpose of this contribution is to present two recent results associated to the phenomenology of proton form factors (FFs) and discuss eventual connections with radiation emission issues. 

Precise data on the proton time-like 
form factor measured by the BABAR collaboration, using the ISR method, show intriguing structures. By plotting these 
data as a function of the 3-momentum of 
the relative motion of the final proton and antiproton, 
a systematic sinusoidal modulation appears in the 
near-threshold region, typical of an interference pattern.  Fourier analysis shows that it may be due to rescattering processes at a relative distance of 0.7-1.5 fm between the centers of the forming hadrons,  (Fig. 1), interfering with the processes at a much smaller scale driven by the quark dynamics \cite{Bianconi:2015owa,Bianconi:2015vva}. 

\begin{figure}[htp!]
\begin{center}
\includegraphics[width=8.cm]{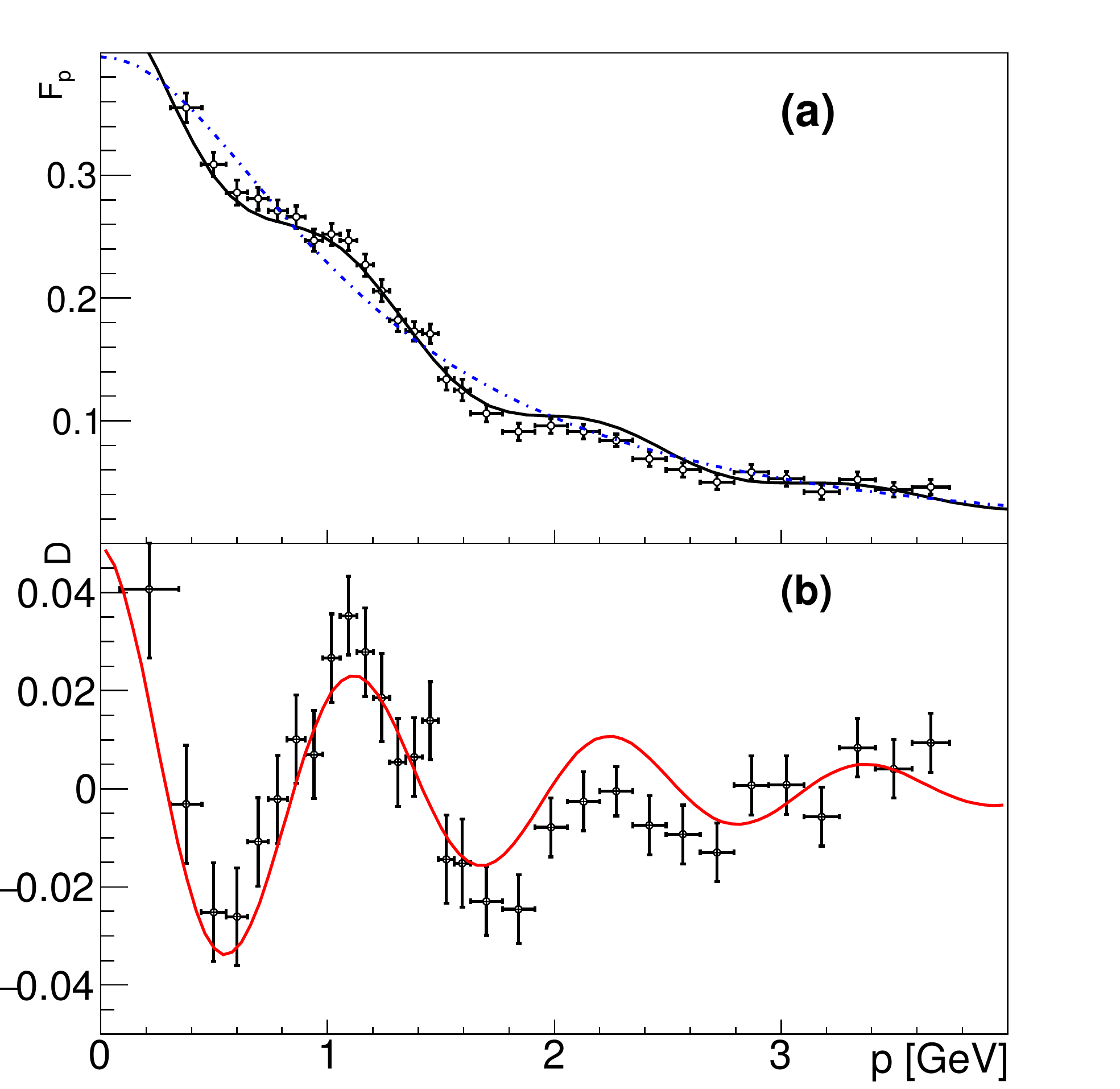}
\caption{ a) FF data from BABAR as a function of $p_{Lab}$, global fit: black solid line. b) after subtraction of a background function (blue dotted line), damped regular fit function (red solid line). }
\label{Fig:lab}
\end{center}
\end{figure}

In the space-like region of transferred momenta, a longstanding issue is the  discrepancy between the extraction of proton measurement of the FF ratio through unpolarized (Rosenbluth method) and polarized (Akhiezer-Rekalo recoil proton polarization method) electron proton scattering. For a recent review, see \cite{Pacetti:2015iqa}. The discrepancy has been attributed to radiative corrections, either precisely recalculated, or including model-dependent two-photon exchange contribution.
A reanalysis of  unpolarized electron-proton elastic scattering data is done in terms of the electric to magnetic form factor squared ratio, that is in principle more robust against experimental correlations and global normalizations. A critical review of the data and of the normalization used in the original analysis show that  the results may  indeed be compatible within the experimental errors and limits are set on the kinematics where the physical information on the form factors can be safely extracted \cite{Pacetti:2016tqi} (Fig. 2).  

The agreement withe the polarized data (black solid circles) can be obtained up to 6 GeV$^2$. Revision of radiative corrections applied to the data at first order or at higher orders using the structure function method may also bring the results into agreement up to 3-4 GeV$^2$, as they have the similar effect to reduce the slope of the Rosenbluth fit.

\begin{figure}[htp!]
\begin{center}
\includegraphics[width=8.cm]{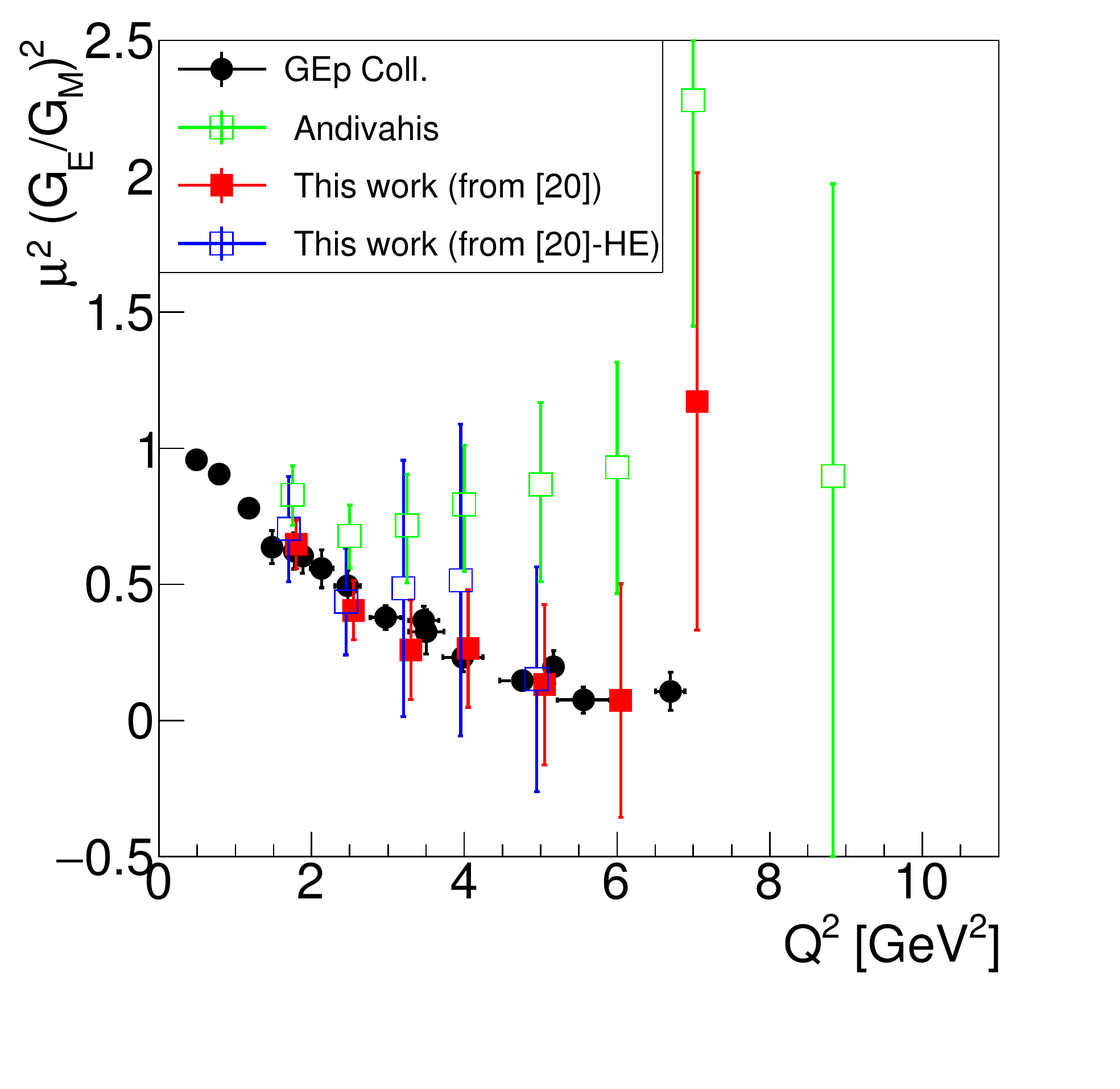}
\caption{ a) FF data from BABAR as a function of $p_{Lab}$, global fit: black solid line. b) after subtraction of a background function (blue dotted line), damped regular fit function (red solid line). }
\label{Fig:lab}
\end{center}
\end{figure}

\newpage

\section{List of participants}

\begin{flushleft}
\begin{itemize}
\item Carlo M. Carloni Calame, University of Pavia, {\tt carlo.carloni.calame@pv.infn.it}
\item Veronica de Leo, INFN Sezione de Roma 3, {\tt veronica.deleo@roma3.infn.it}
\item Simon Eidelman,  Budker Institute of Nuclear Physics, Novosibirsk State University, {\tt eidelman@mail.cern.ch}
\item Fedor Ignatov, Budker Institute of Nuclear Physics, {\tt F.V.Ignatov@inp.nsk.su }
\item Vyacheslav Ivanov, Budker Institute of Nuclear Physics, ${\tt vyacheslav_lvovich_ivanov@mail.ru}$
\item Alex Keshavarzi, University of Liverpool, {\tt A.I.Keshavarzi@liverpool.ac.uk}
\item Andrzej Kupsc, University of Uppsala, {\tt Andrzej.Kupsc@physics.uu.se}
\item Valery Lyubovitskij, T\"ubingen University and Tomsk Polytechnic University, {\tt valeri.lyubovitskij@uni-tuebingen.de}
\item Pere Masjuan, Universt\"at Mainz, {\tt masjuan@kph.uni-mainz.de}
\item Andreas Nyffeler, Universt\"at Mainz, {\tt nyffeler@kph.uni-mainz.de}
\item Giulia Pancheri, INFN Frascati Nat. Labs, {\tt Giulia.Pancheri@lnf.infn.it }
\item Thomas Teubner, University of Liverpool, {\tt thomas.teubner@liverpool.ac.uk}
\item Egle Tomasi-Gustafsson, CEA Saclay, {\tt egle.tomasi@cea.fr}
\item Szymon Tracz, University of Silesia, {\tt stracz@us.edu.pl}
\item Graziano Venanzoni, LNF, {\tt Graziano.Venanzoni@lnf.infn.it}
\item Ping Wang, Institute of High-energy Physics, Beijing, {\tt wangp@ihep.ac.cn}
\end{itemize}
\end{flushleft}


\begin{thebibliography}{99}
\bibitem{Actis:2010gg1}
  S.~Actis {\it et al.}  [Working Group on Radiative Corrections and Monte Carlo Generators for Low Energies Collaboration],
  Eur.\ Phys.\ J.\ C {\bf 66} (2010) 585
  [arXiv:0912.0749 [hep-ph]].
\bibitem{Czyz:2013zga}
  P.~Masjuan 
  {\it et al.},
  arXiv:1306.2045 [hep-ph].
 \bibitem{Czyz:2013sga}
  H.~Czy\.z 
   {\it et al.},
  arXiv:1312.0454 [hep-ph].
\bibitem{vanderBij:2014mxa}
  J.~J.~van der Bij 
  {\it et al.},
  arXiv:1406.4639 [hep-ph].
\bibitem{Carloni:2014mpa}
  C.~M.~Carloni {\it et al.},
  arXiv:1412.7714 [hep-ph].
\bibitem{Czyz:2015nna}
  H.~Czy\.z {\it et al.},
  arXiv:1507.05768 [hep-ph].
 \end{thebibliography}

\begin{thebibliography}{99}
\bibitem{Czerwinski:2012ry}
E. Czerwinski et al., arXiv:1207.6556.
\bibitem{Agashe:2014kda}
K.A. Olive et al. (Particle Data Group), Chin. Phys. C {\bf 38} 
(2014) 090001 and 2015 update.
\bibitem{Abouzaid:2006kk} 
E. Abouzaid et al. (KTEV Collab.), Phys. Rev. D {\bf 75} (2007) 012004.
\bibitem{Agakishiev:2013fwl}
G. Agakishiev et al. (HADES Collab.), Phys. Lett. B {\bf 731} (2014) 265.
\bibitem{Abegg:1994wx}
R. Abegg et al., Phys. Rev. D {\bf 50} (1994) 92.
\bibitem{Achasov:2015mek}
M.N. Achasov et al. (SND Collab.), Phys. Rev. D {\bf 91} (2015) 092010.
\bibitem{Akhmetshin:2014hxv} 
R.R. Akhmetshin et al. (CMD-3 Collab.), Phys. Lett. B {\bf 740} (2015) 273.
\bibitem{Achasov:2015ark}  
 M.N. Achasov et al. (SND Collab.), JETP Lett. {\bf 102} (2015) 266.
\bibitem{Beddall:2008zza}
A. Beddall and A. Beddall, Eur. Phys. J. C {\bf 54} (2008) 365.
\bibitem{Berghauser:2011zz}
H. Berghauser et al. (A2 Collab.), Phys. Lett. B {\bf 701} (2011) 562.
\bibitem{Dzhelyadin:1980kh}
R. Dzhelyadin et al. (Lepton-G Collab.), Phys. Lett. {\bf 94} (1980) 548.
\bibitem{Ablikim:2015wnx}
M. Ablikim et al. (BES Collab.), Phys. Rev. D {\bf 92} (2015) 012001.
\bibitem{Dzhelyadin:1980ki}
V.A. Viktorov et al. (Lepton-G Collab.), Sov. J. Nucl. Phys.  
{\bf 320} (1980) 520.
\bibitem{Abouzaid:2008cd}
E. Abouzaid et al. (KTEV Collab.), Phys. Rev. Lett. {\bf 100} (2008) 182001.
\bibitem{KLOE2:2011aa}
F. Ambrosino et al. (KLOE Collab.),  Phys. Lett. B {\bf 702} (2011) 324.  
\bibitem{Berlowski:2007aa}
M. Berlowski et al. (WASA Collab.), Phys. Rev. D {\bf 77} (2008) 032004.
\bibitem{Ambrosino:2008cp}
F. Ambrosino et al. (KLOE Collab.),  Phys. Lett. B {\bf 675} (2009) 283.  
\bibitem{Ablikim:2013wfg}
M. Ablikim et al. (BES Collab.),  Phys. Rev. D {\bf 87} (2013) 092011.  
\bibitem{Babusci:2012ft1}
D. Babusci et al. (KLOE Collab.),  Phys. Lett. B {\bf 718} (2013) 910. 
\bibitem{Pedlar:2009aa}
T.K. Pedlar et al. (CLEO Collab.),  Phys. Rev. D {\bf 79} (2009) 111101.
\bibitem{Ablikim:2015eos}
M. Ablikim et al. (BES Collab.), Phys. Rev. D {\bf 92} (2015) 051101.  

\end{thebibliography}

\begin{thebibliography}{99}
\bibitem{Stollenwerk:2011zz}
  F.~Stollenwerk, C.~Hanhart, A.~Kupsc, U.~G.~Mei{\ss}ner and A.~Wirzba,
  Phys.\ Lett.\ B {\bf 707} (2012) 184.
\bibitem{Hanhart:2013vba}
  C.~Hanhart, A.~Kupsc, U.-G.~Mei{\ss}ner, F.~Stollenwerk and A.~Wirzba,
  Eur.\ Phys.\ J.\ C {\bf 73} (2013) 2668
   Erratum: [Eur.\ Phys.\ J.\ C {\bf 75} (2015) 242].
\bibitem{Kubis:2015sga}
  B.~Kubis and J.~Plenter,
  Eur.\ Phys.\ J.\ C {\bf 75} (2015) 283.
\bibitem{Adlarson:2011xb}
  P.~Adlarson {\it et al.} [WASA-at-COSY Collaboration],
  Phys.\ Lett.\ B {\bf 707} (2012) 243.
\bibitem{Babusci:2012ft}
  D.~Babusci {\it et al.} [KLOE Collaboration],
  Phys.\ Lett.\ B {\bf 718} (2013) 910.
\bibitem{Hoferichter:2014vra}
  M.~Hoferichter, B.~Kubis, S.~Leupold, F.~Niecknig and S.~P.~Schneider,
  Eur.\ Phys.\ J.\ C {\bf 74} (2014) 3180.
\bibitem{Xiao:2015uva}
  C.~W.~Xiao, T.~Dato, C.~Hanhart, B.~Kubis, U.-G.~Mei{\ss}ner and A.~Wirzba,
  arXiv:1509.02194 [hep-ph].

\end{thebibliography}

\begin{thebibliography}{99}

\bibitem{JN_09} 
  F.~Jegerlehner and A.~Nyffeler,
  Phys.\ Rept.\  {\bf 477}, 1 (2009). 

\bibitem{g-2_reviews_exp} 
  J.~P.~Miller {\it et al.}, 
  Ann.\ Rev.\ Nucl.\ Part.\ Sci.\  {\bf 62}, 237 (2012); 
  K.~A.~Olive {\it et al.} [Particle Data Group Collaboration],
  Chin.\ Phys.\ C {\bf 38}, 090001 (2014); 
  G.~W.~Bennett {\it et al.},  [Muon g-2 Collaboration],
  Phys.\ Rev.\ D {\bf 73}, 072003 (2006). 


\bibitem{g-2_hadronic}
  T.~Blum {\it et al.}, 
  arXiv:1311.2198 [hep-ph]; 
  M.~Benayoun {\it et al.},
  arXiv:1407.4021 [hep-ph].

\bibitem{future_g-2_exp}
  D.~W.~Hertzog,
  EPJ Web Conf.\  {\bf 118}, 01015 (2016). 

\bibitem{HLbL_review}
  J.~Bijnens,
  EPJ Web Conf.\  {\bf 118}, 01002 (2016). 

\bibitem{HLbL_Lattice}
  T.~Blum {\it et al.}, 
  Phys.\ Rev.\ Lett.\  {\bf 114}, 012001 (2015); 
  T.~Blum {\it et al.}, 
  Phys.\ Rev.\ D {\bf 93}, 014503 (2016);  
  J.~Green {\it et al.}, 
  Phys.\ Rev.\ Lett.\  {\bf 115}, 222003 (2015); 
  J.~Green {\it et al.}, 
  arXiv:1510.08384 [hep-lat].


\bibitem{HLbL_DR} 
  G.~Colangelo {\it et al.}, 
  JHEP {\bf 1409}, 091 (2014); 
  G.~Colangelo {\it et al.}, 
  Phys.\ Lett.\ B {\bf 738}, 6 (2014); 
  G.~Colangelo {\it et al.}, 
  JHEP {\bf 1509}, 074 (2015); 
%
  G.~Colangelo, talk at this workshop; 
%
  V.~Pauk and M.~Vanderhaeghen,
  arXiv:1403.7503 [hep-ph]; 
%
  V.~Pauk and M.~Vanderhaeghen,
  Phys.\ Rev.\ D {\bf 90}, 113012 (2014). 


\bibitem{Nyffeler_16}
  A.~Nyffeler,
  arXiv:1602.03398 [hep-ph].


\bibitem{LMDV} 
  M.~Knecht and A.~Nyffeler,
  Eur.\ Phys.\ J.\ C {\bf 21}, 659 (2001); 
%
  Phys.\ Rev.\ D {\bf 65}, 073034 (2002). 


\bibitem{BESIII_single_virtual}
A.~Denig [BESIII Collaboration],
  Nucl.\ Part.\ Phys.\ Proc.\  {\bf 260}, 79 (2015). 


\bibitem{DR_pion_TFF}
  M.~Hoferichter {\it et al.}, 
  Eur.\ Phys.\ J.\ C {\bf 74}, 3180 (2014). 

\bibitem{PdeRV_09}
  J.~Prades, E.~de Rafael and A.~Vainshtein,
  Adv.\ Ser.\ Direct.\ High Energy Phys.\  {\bf 20}, 303 (2009) 
  [arXiv:0901.0306 [hep-ph]].


\end{thebibliography}

\begin{thebibliography}{99}

\bibitem{AdSQCD}
  T.~Gutsche, V.~E.~Lyubovitskij, I.~Schmidt, A.~Vega,
  Phys. Rev. D {\bf 85}, 076003 (2012);
  Phys. Rev. D {\bf 86}, 036007 (2012);
  Phys. Rev. D {\bf 87}, 016017 (2013);
  Phys. Rev. D {\bf 87}, 056001 (2013);
  A.~Vega {\it et al.}, 
  Phys. Rev. D {\bf 80}, 055014 (2009);
  T.~Branz {\it et al.}, 
  Phys. Rev. D {\bf 82}, 074022 (2010).

\bibitem{Gutsche:2015xva}
  T.~Gutsche, V.~E.~Lyubovitskij, I.~Schmidt, A.~Vega,
  Phys. Rev. D {\bf 91}, 114001 (2015).

  \bibitem{Brodsky:1983vf}
  S.~J.~Brodsky {\it et al.}, 
  Phys. Rev. Lett. {\bf 51}, 83 (1983);
  C.~E.~Carlson, F.~Gross,
  Phys. Rev. Lett. {\bf 53}, 127 (1984);  
  S.~J.~Brodsky, J.~R.~Hiller,
   Phys. Rev. D {\bf 46}, 2141 (1992).

\end{thebibliography}

\begin{thebibliography}{99}

\bibitem{Masjuan:2015lca}
  P.~Masjuan and P.~Sanchez-Puertas,
  arXiv:1504.07001 [hep-ph];
  P.~Masjuan and P.~Sanchez-Puertas,
  arXiv:1512.09292 [hep-ph].

\bibitem{Masjuan:2008fv}
  P.~Masjuan, S.~Peris and J.~J.~Sanz-Cillero,
  Phys.\ Rev.\ D {\bf 78} (2008) 074028
  [arXiv:0807.4893 [hep-ph]].

\bibitem{Masjuan:2012wy}
  P.~Masjuan,
  Phys.\ Rev.\ D {\bf 86} (2012) 094021
  [arXiv:1206.2549 [hep-ph]].
  
\bibitem{Escribano:2013kba}
  R.~Escribano, P.~Masjuan and P.~Sanchez-Puertas,
  Phys.\ Rev.\ D {\bf 89} (2014) no.3,  034014
  [arXiv:1307.2061 [hep-ph]].

\bibitem{Escribano:2015nra}
  R.~Escribano, P.~Masjuan and P.~Sanchez-Puertas,
  Eur.\ Phys.\ J.\ C {\bf 75} (2015) no.9,  414
  [arXiv:1504.07742 [hep-ph]].

\bibitem{Escribano:2015vjz}
  R.~Escribano and S.~Gonzˆlez-Sol's,
  arXiv:1511.04916 [hep-ph].

\bibitem{Escribano:2015yup}
  R.~Escribano, S.~Gonzalez-Solis, P.~Masjuan and P.~Sanchez-Puertas,
  arXiv:1512.07520 [hep-ph].

 \bibitem{Baker}
  G.~A.~Baker and P.~Graves-Morris, Encyclopedia of Mathematics and its Applications, Cambridge Univ. Press, 1996;   P.~Masjuan,
  arXiv:1005.5683 [hep-ph].

 \bibitem{Chisholm}
J.~S.~R. Chisholm,
Mathematics of Computation, {\bf27} (124), 1973;
J.~S.~R. Chisholm and J.~McEwan.
Proc. R. Soc. Lond. A, {\bf 336} 421-452, 1974.


\bibitem{Masjuan:2007ay}
  P.~Masjuan and S.~Peris,
  JHEP {\bf 0705} (2007) 040
  [arXiv:0704.1247 [hep-ph]];
  Phys.\ Lett.\ B {\bf 663} (2008) 61
  [arXiv:0801.3558 [hep-ph]].

\bibitem{Aguar-Bartolome:2013vpw}
  P.~Aguar-Bartolome {\it et al.} [A2 Collaboration],
  Phys.\ Rev.\ C {\bf 89} (2014) no.4,  044608
  [arXiv:1309.5648 [hep-ex]].
  
  \bibitem{Queralt:2010sv}
  P.~Masjuan Queralt,
  arXiv:1005.5683 [hep-ph].
  
  \bibitem{Arnaldi:2016pzu}
  R.~Arnaldi {\it et al.} [NA60 Collaboration],
  Phys.\ Lett.\ B {\bf 757} (2016) 437.
  
  \bibitem{inprep}
  P.~Masjuan and P.~Sanchez-Puertas, in preparation.
 
  \bibitem{pi0data}
  H.~J.~Behrend {\it et al.} [CELLO Collaboration],
  Z.\ Phys.\ C {\bf 49} (1991) 401;
  J.~Gronberg {\it et al.} [CLEO Collaboration],
  Phys.\ Rev.\ D {\bf 57} (1998) 33
  [hep-ex/9707031];
  B.~Aubert {\it et al.} [BaBar Collaboration],
  Phys.\ Rev.\ D {\bf 80} (2009) 052002
  [arXiv:0905.4778 [hep-ex]].
  S.~Uehara {\it et al.} [Belle Collaboration],
  Phys.\ Rev.\ D {\bf 86} (2012) 092007
  [arXiv:1205.3249 [hep-ex]].
  
\bibitem{Landsberg:1986fd}
  L.~G.~Landsberg,
  Phys.\ Rept.\  {\bf 128} (1985) 301.

\bibitem{Masjuan:2014rea}
  P.~Masjuan,
  Nucl.\ Part.\ Phys.\ Proc.\  {\bf 260} (2015) 111
  [arXiv:1411.6397 [hep-ph]].

\end{thebibliography}

\vspace{-0.9cm}
\begin{thebibliography}{99}

\bibitem{amaldi}E. Amaldi, {\it The Bruno Touschek Legacy},   1981, CERN 81-19

\bibitem{bernardini}C. Bernardini, {\it Ada: the first electron-positron collider},  
Phys. in Persp., {\bf 6} (2004) 156-183 
\bibitem{bonolis} L. Bonolis, {\it    Bruno Touschek vs Machine Builders. AdA, the first matter-antimatter collider} , 
La Rivista del Nuovo Cimento, {\bf 28} issue 11 (2005)  1-60

\bibitem{bonolispancheri}L. Bonolis  and G. Pancheri,  {\it Bruno Touschek: particle physicist and father of the $ e^+e^-$  collider},  European Physical  Journal H  {\bf 36}, issue 1 (2011) 1-61 

\bibitem{grecopancheri}M. Greco and G. Pancheri (eds.). 2005. {\it Bruno Touschek Memorial Lectures},   
  http://www.lnf.infn.it/sis/frascatiseries/Volume33/volume33.pdf

\bibitem{bernardinipancheripellegrini} C. Bernardini,  G. Pancheri and C. Pellegrini, {\it  Bruno Touschek, from Betatrons to Electron-positron Colliders.}  Rev.  Acc. Science and Technology {\bf 1} (2015)  1-23

\end{thebibliography}

\vspace{-0.5cm}

\begin{thebibliography}{99}

\bibitem{Hagiwara:2011af}
  K.~Hagiwara, R.~Liao, A.~D.~Martin, D.~Nomura and T.~Teubner,
  J.\ Phys.\ G {\bf 38} (2011): 085003

\bibitem{Aulchenko:2001je} 
  V.~M.~Aulchenko {\it et al.},
  BUDKER-INP-2001-45

\bibitem{Khazin:2008zz}
  B.~Khazin, 
  Nucl.\ Phys.\ Proc.\ Suppl.  {\bf 181-182} (2008) 376
\bibitem{Shatunov:2000zc}
  Y.~.M.~Shatunov {\it et al.},
  Conf.\ Proc.\ C {\bf 0006262} (2000): 439
\bibitem{Berkaev:2012qe}
  D.~Berkaev {\it et al.},
  Nucl.\ Phys.\ Proc.\ Suppl.\  {\bf 225-227} (2012): 303

\bibitem{Actis:2010gg}
  S.~Actis {\it et al.}, 
  Eur.\ Phys.\ J.\ C {\bf 66} (2010): 585

\bibitem{Arbuzov:2005pt}
  A.~B.~Arbuzov, G.~V.~Fedotovich, F.~V.~Ignatov, E.~A.~Kuraev and A.~L.~Sibidanov,
  Eur.\ Phys.\ J.\ C {\bf 46} (2006): 689

\bibitem{Balossini:2006wc}
  G.~Balossini, C.~M.~Carloni Calame, G.~Montagna, O.~Nicrosini and F.~Piccinini,
  Nucl.\ Phys.\ B {\bf 758} (2006): 227

\end{thebibliography}

\begin{thebibliography}{99}

\bibitem{Hagiwara:2011af}
  K.~Hagiwara, R.~Liao, A.~D.~Martin, D.~Nomura and T.~Teubner,
  J.\ Phys.\ G {\bf 38} (2011) 085003
  [arXiv:1105.3149 [hep-ph]].
  
\bibitem{Bennett:2006fi}
  G.~W.~Bennett {\it et al.} [Muon g-2 Collaboration],
  Phys.\ Rev.\ D {\bf 73} (2006) 072003
  [hep-ex/0602035].
  
\bibitem{Grange:2015fou}
  J.~Grange {\it et al.} [Muon g-2 Collaboration],
 [arXiv:1501.06858 [hep-ex]].
  
\bibitem{Iinuma:2011zz}
  H.~Iinuma [J-PARC muon g-2/EDM Collaboration],
  J.\ Phys.\ Conf.\ Ser.\  {\bf 295} (2011) 012032.
  
\bibitem{Ball:2009qv}
  R.~D.~Ball {\it et al.} [NNPDF Collaboration],
  JHEP {\bf 1005} (2010) 075
  [arXiv:0912.2276 [hep-ph]].
  
\bibitem{Benayoun:2015gxa}
  M.~Benayoun, P.~David, L.~DelBuono and F.~Jegerlehner,
  arXiv:1507.02943 [hep-ph].
  
\bibitem{Aubert:2009ad}
  B.~Aubert {\it et al.} [BaBar Collaboration],
  Phys.\ Rev.\ Lett.\  {\bf 103} (2009) 231801
  [arXiv:0908.3589 [hep-ex]].
  
\bibitem{Hagiwara:2003da}
  K.~Hagiwara, A.~D.~Martin, D.~Nomura and T.~Teubner,
  Phys.\ Rev.\ D {\bf 69} (2004) 093003
[hep-ph/0312250];
  Phys.\ Lett.\ B {\bf 649} (2007) 173
  [hep-ph/0611102].
  
\bibitem{Babusci:2012rp}
  D.~Babusci {\it et al.} [KLOE Collaboration],
  Phys.\ Lett.\ B {\bf 720} (2013) 336
  [arXiv:1212.4524 [hep-ex]].
  
\bibitem{DeLeo:2015yja}
  V.~De Leo [KLOE KLOE-2 Collaboration],
  Acta Phys.\ Polon.\ B {\bf 46} (2015) 45
  [arXiv:1501.04446 [hep-ex]].
  
\bibitem{Ablikim:2015orh}
  M.~Ablikim {\it et al.} [BESIII Collaboration],
  Phys.\ Lett.\ B {\bf 753} (2016) 629
  [arXiv:1507.08188 [hep-ex]].

\end{thebibliography}

\begin{thebibliography}{99}
\bibitem{Fred_2003} F. Jegerlehner  J. Phys. G: Nucl. Part. Phys. 29 (2003) 101 
\bibitem{Aoyama} T. Aoyama, M. Hayakawa, T. Kinoshita, M. Nio, Phys. Rev. D77, 053012 (2008), 0712.2607.
\bibitem{passera}  M.~Steinhauser,
 Phys.\ Lett.\ B {\bf 429} (1998) 158.\\
\bibitem{kloe2} A.~Anastasi {\it et al.}, KLOE-2 Coll., 
{\it to be submitted to Phys. Lett. B}.
\bibitem{PHOKHARA} 
H. Czy\.{z} , A. Grzelinska, J.H. K\"uhn, G. Rodrigo, Eur. Phys. J. C 39 (2005) 411. 
\bibitem{Venanzo}
D. Babusci et al. Phy.Lett. B 720 336-343 (2013).

\end{thebibliography}

\begin{thebibliography}{99}
\bibitem{Bianconi:2015owa} 
  A.~Bianconi and E.~Tomasi-Gustafsson,
  Phys.\ Rev.\ Lett.\  {\bf 114}, 232301 (2015).
  
\bibitem{Bianconi:2015vva} 
  A.~Bianconi and E.~Tomasi-Gustafsson,
  Phys.\ Rev.\ C {\bf 93}, 035201 (2016).
  
\bibitem{Pacetti:2015iqa} 
  S.~Pacetti, R.~Baldini Ferroli and E.~Tomasi-Gustafsson,
  Phys.\ Rept.\  {\bf 550-551}, 1 (2015).
\bibitem{Pacetti:2016tqi} 
  S.~Pacetti and E.~T.~Gustafsson,
  arXiv:1604.02421 [nucl-th].

\end{thebibliography}
\end{document}